\documentclass[rpreprint,3p,10pt,sort&compress, colorlinks,linkcolor=blue, citecolor=blue]{elsarticlex}

\usepackage[utf8]{inputenc}

\usepackage{amssymb}
\usepackage{amsmath,mathrsfs}
\usepackage{color}

\usepackage{siunitx}

\usepackage{stfloats}

\usepackage{bm}

\usepackage{graphicx}

\usepackage{caption}
\usepackage{subcaption}

\usepackage{multicol}

\usepackage{multirow}

\usepackage{mhchem}

\usepackage[toc,page]{appendix}

\usepackage{bm}

\usepackage{float}

\usepackage[colorlinks,linkcolor=blue,anchorcolor=green,citecolor=blue]{hyperref}

\begin{document}
	
	\begin{frontmatter}
		\title{Three-dimensional non-isothermal phase-field modeling of microstructure evolution during selective laser sintering}
		\author[els]{Yangyiwei Yang}
		
		\author[rvt]{Olav Ragnvaldsen}
		
		\author[els]{Yang Bai}
		
		\author[els]{Min Yi\corref{cor1}}
		\ead{yi@mfm.tu-darmstadt.de}
		
		\author[els]{Bai-Xiang Xu\corref{cor1}}
		\ead{xu@mfm.tu-darmstadt.de}
		
		\cortext[cor1]{Corresponding author}
		
		\address[els]{Mechanics of Functional Materials Division, Institute of Materials Science, Technische Universit\"at Darmstadt, Darmstadt 64287, Germany}
		
		\address[rvt]{Department of Materials Science and Engineering, Norwegian University of Science and Technology, 7491 Trondheim, Norway}

		
		\begin{abstract}
			Predicting the microstructure during selective laser sintering (SLS) is of great interests, which can compliment the current time and cost expensive trial-and-error principle with an efficient computational design tool. However, it still remains a great challenge to simulate the microstructure evolution during SLS due to the complex underlying physical phenomena. 
			In this work, we present a three-dimensional finite element phase-field simulation of the SLS single scan, and revealed the process-microstructure relation. We use a thermodynamically consistent non-isothermal phase-field model including various physics (i.e. partial melting, pore structure evolution, diffusion, grain boundary migration, and coupled heat transfer), and interaction of powder bed and laser power absorption. The initial powder bed is generated by the discrete element method. Moreover, we present in the manuscript a novel algorithm analogy to minimum coloring problem and managed to simulate a system of 200 grains with grain tracking using as low as 8 non-conserved order parameters. The developed model is shown to capture interesting phenomena which are not accessible to the conventional isothermal model. Specifically, applying the model to SLS of the stainless steel 316L powder, we identify the influences of laser power and scanning speed on microstructural indicators, including the porosity, surface morphology, temperature profile, grain geometry, and densification. We further validate the first-order kinetics during the porosity evolution, and demonstrate the applicability of the developed model in predicting the linkage of densification factor to the specific energy input during SLS.
		\end{abstract}
		
		
	\end{frontmatter}
	
	\section{Introduction}\label{Introduction}
	Selective laser sintering (SLS) is a typical additive manufacturing process meant for rapid prototyping and tooling \cite{kumar2003selective, kruth1998progress, khaing2001direct, karapatis1998direct, gu2012laser}. During SLS, a desired geometry is built by sequentially layer-by-layer powder spreading and subsequent laser scanning, whereby the photonic energy is transformed into heat by absorption \citep{Gusarov2009,gusarov2005modelling}. To be distinguished with the other laser-based powder bed additive manufacturing known as selective laser melting (SLM), there is no significant melting phenomenon during SLS. Whereas the temperature is sufficiently high for particles to bind together through sorts of mechanisms, leading to the products with relatively high porosity \cite{kruth2005binding,gu2012laser, Li2018}. Because of these characteristics, SLS has been applied for the industrial production of individually designed components made of organic polymers, ceramics, and metallic alloys which have relatively high melting or transition temperature \cite{sing2017direct,kumar2003selective,ko2007all,gu2012laser}. It also shows the possibility to produce the porous biomaterials, especially medical scaffold and bones \cite{williams2005bone,eshraghi2010mechanical,tan2005selective,berry1997preliminary}, and functional materials \cite{sing2017direct,guo2004rapid,jhong2016fabricating}.
	
	Although the procedure of SLS is conceptually simple, the underlying physics is complex and covers a broad range of time and length scales. Since its birth \citep{deckard1987recent}, identification and comprehension of those phenomena during SLS and their effects on the final product are crucial for the successful manufacturing, which highly relies on trial-and-error and even sort of empirical knowledge \cite{gu2012laser,simchi2006direct, olakanmi2015review}. There are already many computational works performed regarding the underlying physics during SLS, where powder-laser interaction \cite{gusarov2005modelling,Gusarov2010,xiao2008numerical,Zohdi2017} and heat transfer \cite{dong2009three,Ganeriwala2014,liu2012micro,arisoy2019modeling} have been massively investigated. Apart from those, the microstructure evolution has also gained great interests since it is directly related to mechanical properties of the final products like tensile strength, ductility, and fracture toughness. Up to now, however, it still remains a great challenge to simulate the microstructure evolution if the influences from all aspects are considered. During scanning, there is drastic difference in the thermal conditions among particles due to different exposure to the laser beam. Some of particles may even partially melted during the process \cite{gu2012laser,agarwala1995direct}. Therefore, binding mechanism for certain particle/grain may vary from the solid-state sintering to the liquid-state sintering, and even melting-solidification according to the intensity of its partial melting \cite{kruth2005binding,agarwala1995direct,olakanmi2013selective,klocke2003coalescence}. For the same reason, very high temperature gradients and cooling rates also exist \cite{keller2017application,gusarov2005modelling}, which make the mechanism of grain coalescence and coarsening during SLS deviate from that in conventional isothermal sintering.  
	
	In recent decades, the phase-field method has been utilized to simulate the microstructure evolution during sintering-related techniques because of its ability to model the complex spatial geometries without explicitly tracking the position of the surface and interface. One of the earliest attempts to combine such efforts into a phase-field sintering model is carried out by Wang \textit{et al.}, who utilized a conserved density field and a set of non-conserved orientation fields \citep{wang1993thermodynamically,kazaryan1999generalized}. Based on this model, some featured details, such as the dihedral angle, neck growth, shrinkage and the grain growth kinetics with porosity, have been captured and validated by experiments \citep{ahmed2013phase,ahmed2014phase,millett2012phase}. In recent research, Zhang \textit{et al.} adopted this isothermal model to investigate the necking among several powders during the SLS \citep{zhang2018phase}, whereas the information of non-isothermal profile, i.e. inhomogeneous temperature field with high temperature gradient, is missing. Lu \textit{et al.} developed a two-dimensional (2D) phase-field model to study the molten pool geometry, porosity, and grain structure during powder bed-based additive manufacturing \citep{lu2018phase}. This model presented a good coherency of porosity and grain structure with the experimental data. However, it is not accessible to the features like the surface morphology and grain geometry due to the limitation of 2D presentation. In brief, a three-dimensional (3D) non-isothermal phase-field model is still in demand to improve the understanding of the microstructure evolution during the SLS process.
	
	In this work, we present a 3D non-isothermal phase-field model to investigate the microstructure evolution during the SLS process. The model is derived in a thermodynamically consistent way according to our latest work in \citep{yang2018arxiv} and numerically implemented with the finite element method (FEM). Besides the diffusion and grain boundary migration mechanisms, the role of the partial melting of the powders is also considered. A scenario with a conserved order parameter (OP) as well as a series of non-conserved ones is utilized to represent the microstructure evolution during SLS. The discrete element method (DEM) is used to generate the initial powder bed. The Welsh-Powell algorithm for the minimum coloring problem (MCP) is also applied to optimize the profile of OPs to reduce the computation cost due to large amount of variables. This model is demonstrated to be capable of capturing interesting phenomena which are not accessible to the conventional isothermal model. We further perform phase-field simulations on the SLS processing of stainless steel 316L powder, in which the temperature-dependent model parameters are readily extracted from the experimental measurement of surface and interface energies. The influences from the laser power and scanning speed on key features, such as the porosity, surface morphology, temperature profile, grain geometry and densification, are also discussed. It is hoped that the present work could enrich the modeling and computational toolkit which is practicable for the simulation of SLS-based additive manufacturing.
	
	
	\section{Modeling}\label{Modelling}

	\subsection{Non-isothermal Phase-field Scenario}
\begin{figure*}[!t]
		\centering
		\includegraphics[width=16cm]{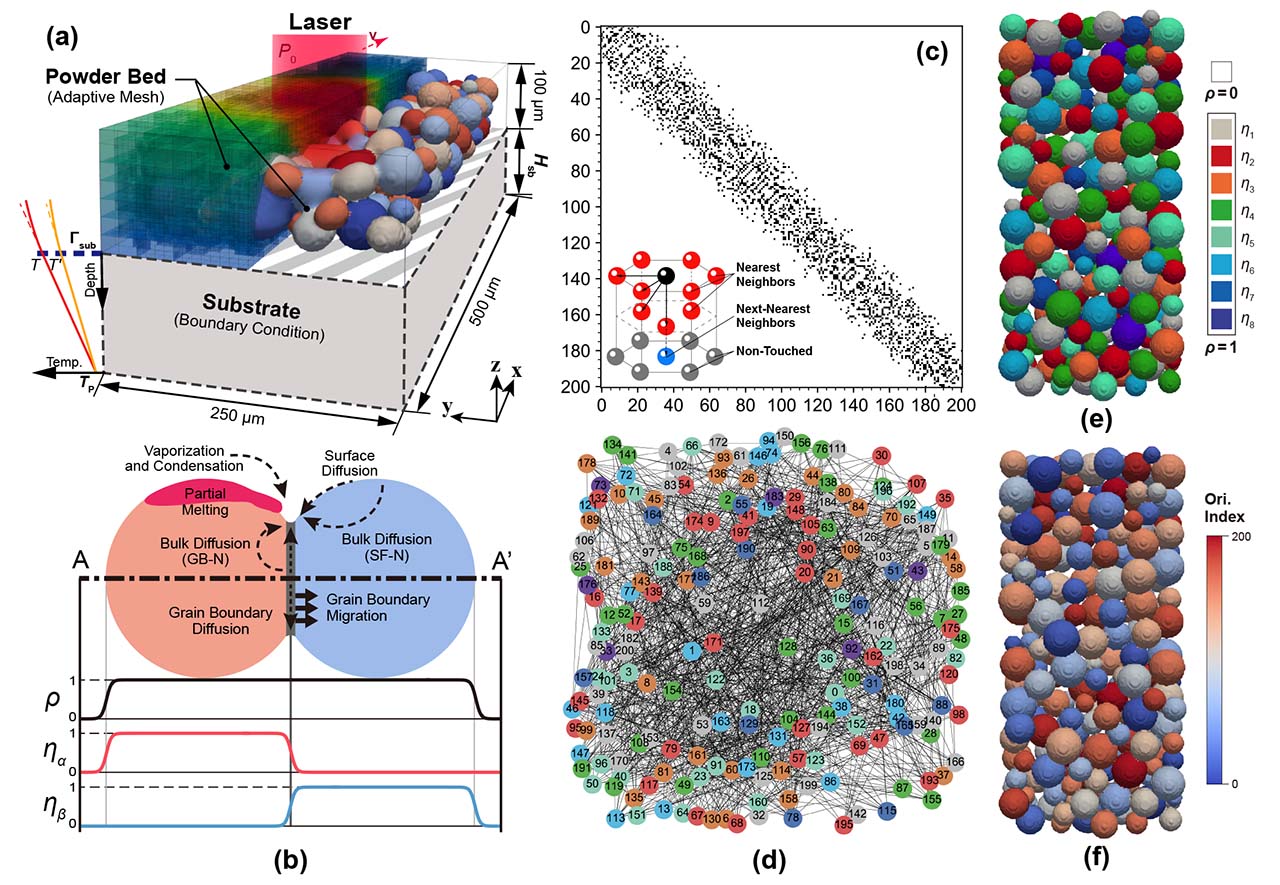}
		\caption{Schematics for the powder bed scenario of the SLS system. (a) Geometry of the powder bed with a thick substrate; (b) An illustration of order parameters ($\rho, \eta_\alpha, \eta_\beta$) profile across A-A' section including physical phenomena, i.e. different diffusion mechanisms and grain boundary migration;  (c) adjacency matrix of the powder bed. Inset: the distance of the next-nearest neighbor in hexagonal close packing of the largest particles as the adjacency criterion; (d) coloring map $G(\mathbb{V},\mathbb{E})$ generated by using the adjacency matrix in (c); (e) profile of the non-conserved OPs; (f) indices of the unique particles in the powder bed.}
		\label{fig1}
	\end{figure*}

	A segment of a powder bed is considered, as shown in Fig. \ref{fig1}a. A domain with the dimension of $250\times500\times100~\si{\micro\meter}^3$ 
	is utilized to simulate a powder bed. Both conserved and non-conserved order parameters (OPs) $\rho$ and $\{\eta_\alpha\}$ are employed to represent the powder bed with multiple particles. The conserved OP $\rho$ indicates the substance while the non-conserved OPs $\eta_\alpha$ distinguish particles with different crystallographic orientations \citep{wang2006computer,ahmed2013phase,yang2018arxiv}. $\rho=1$ and $\rho=0$ represent the substance and atmosphere/pore region, respectively. In each grain within the substance, only one of $\eta_\alpha$ takes unity and others are zero (Fig. \ref{fig1}b). These grains and grain boundaries have $\rho=1$ (we assume the density variation across the grain boundary is negligible). When $\rho=0$, no grain is present. This profile of OPs leads to the constraint $(1-\rho)+\sum_{\alpha} \eta_\alpha=1$. Their temporal evolution indicates the changes of the surface and grain boundaries, and thus represents the microstructure evolution during the SLS process. 
	
	The non-isothermal formulations following our latest work in Ref. \citep{yang2018arxiv} are extended with the contribution from partial melting. The total free energy density  $f(T,\rho,\{\eta_\alpha\})$, derived through the Legendre transform of the internal energy and entropy \citep{penrose1990thermodynamically}, is formulated as
	\begin{equation}
	\begin{split}
	f(T,\rho,\{\eta_\alpha\})=& \xi f_\text{ht}(T)\left(\underline{A}\rho + \underline{B}\sum_\alpha\eta_\alpha\right) +
	\underline{C}\left[\rho^2(1-\rho)^2 \right] + \\
	&\underline{D}\left[\rho^2+6(1-\rho)\sum_\alpha\eta_\alpha^2 - 4(2-\rho)\sum_\alpha\eta_\alpha^3 + 3\left(\sum_\alpha\eta_\alpha^2 \right)^2 \right] + \\
	&\frac{1}{2} T \kappa_\rho \left| \nabla \rho \right|^2 + \frac{1}{2} T \kappa_\eta \sum_\alpha \left| \nabla \eta_\alpha \right|^2,
	\label{eqm2}
	\end{split}
	\end{equation}
	where
	\begin{equation*}
	\begin{split}
	& f_\text{ht}(T)=T\int e_\text{ht}(T)\text{d}\left(\frac{1}{T}\right), \\
	& \underline{C}=\underline{C}_\text{pt} - \underline{C}_\text{cf}(T-T_\text{M}) , \\
	& \underline{D}=\underline{D}_\text{pt} - \underline{D}_\text{cf}(T-T_\text{M}).
	\end{split}
	\end{equation*}
	The eight model parameters
	($\underline{A},\underline{B},\underline{C}_\text{pt}, \underline{C}_\text{cf}, \underline{D}_\text{pt}, \underline{D}_\text{cf},~\kappa_\eta$ and $\kappa_\rho$) in Eq. (\ref{eqm2}) are obtained from the experimentally measured temperature-dependent surface and grain-boundary energies ($\gamma_\text{sf}$ and $\gamma_\text{gb}$) and the width of the grain boundary. Fitting coefficient $\xi$ is utilized to help fitting with the experimental result (see Supplementary Information). $e_\text{ht}$ is the gain (or loss) of the internal energy density due to the heat transfer, which can be formulated as (note that we have set the internal energy of the atmosphere/pores at melting point $T_\text{M}$ of the system as zero)
	\begin{equation}
	e_\text{ht}(T)=(c_\text{bk}-c_\text{at})(T-T_\text{M}) +\mathscr{L}\Phi(\tau).
	\label{eqm3}
	\end{equation}
	In Eq. (\ref{eqm3}), $c$ is the volumetric specific heat. The subscript `bk' and `at' denote the substance materials and atmosphere/pores, respectively. The latent heat during the solid-liquid phase transition is simply represented by $\mathscr{L}\Phi(\tau)$, where the sufficiently smooth interpolation function $\Phi(\tau)$ takes one when $\tau =T/T_\text{M}\rightarrow1$ and to zero when $\tau\rightarrow0$. 
	
	The temporal evolution of $\rho$ is derived from the mass conservation, i.e.
	\begin{equation}
	\dot{\rho }+\nabla \cdot {{\mathbf{j}}_\text{diff}}+\nabla \cdot {{\mathbf{j}}_\text{melt}}=0,
	\label{eqmn4}
	\end{equation}
	where ${{\mathbf{j}}_\text{diff}}$ and ${{\mathbf{j}}_\text{melt}}$ represent the mass flux contribution from the diffusion and partial melting, respectively. Here we simply assume that the partial melting is locally restricted to individual particles, and the melt flow is driven by the capillary pressure. The driven force of the melts is described by change of surfacial curvature $A_\text{sf}$ and energy $\gamma_\text{sf}$, i.e. ${{\mathbf{j}}_\text{melt}} \sim -\Phi(\tau)\mathbf{M}_\text{melt} \cdot \nabla(A_\text{sf}\gamma_\text{sf})$. According to Young-Laplace equation in which the on-site chemical potential is proportional to $A_\text{sf}\gamma_\text{sf}$, we can thereby formulate ${{\mathbf{j}}_\text{melt}}$ into a diffusion-like form. ${{\mathbf{j}}_\text{diff}}$ can be obtained directly from Fick's law. In this way, Eq. (\ref{eqmn4}) can be rearranged into the form of the Cahn-Hilliard equation 
	\begin{equation}
	\dot{\rho}(\mathbf{r},t)=
	\nabla \cdot \left[ \mathbf{M} \cdot \nabla \left(\frac{\partial f}{\partial\rho}-T\kappa_\rho\nabla^2\rho \right) \right].
	\label{eqm4}
	\end{equation}
	Here the mobility tensor $\mathbf{M}$ is formulated in the fashion to consider contributions not only from mass transfer paths through bulk (bk), atmosphere (at), surface (sf) and grain boundary (gb) \citep{wang2006computer,asp2006phase,gugenberger2008comparison,ahmed2013phase}, but also from the gain due to the partial melting, i.e.
	\begin{equation}
	\mathbf{M}=\left[\mathbf{M}_\text{bk}+\mathbf{M}_\text{at}+\mathbf{M}_\text{sf}+\mathbf{M}_\text{gb}\right]_\text{diff}+\Phi(\tau)\mathbf{M}_\text{melt}.
	\label{eqmM}
	\end{equation}
	In this regard the contribution of the partial melting is treated more like an enhanced surface diffusion when $\tau \rightarrow1$. To represent such contribution more properly, a ${{\mathbf{j}}_\text{melt}}$ obtained by coupling with fluid dynamic equations (e.g. the Navier-Stokes Equations) should be worked out in the future.
	
	On the other hand, evolution of $\{\eta_\alpha\}$ is governed by Allen-Cahn equation with the corresponding mobility tensor $\mathbf{L}$, i.e.
	\begin{equation}
	\dot{\eta}_\alpha(\mathbf{r},t)= -\mathbf{L} \left(\frac{\partial f}{\partial\eta_\alpha}-T\kappa_\eta\nabla^2\eta_\alpha \right).
	\label{eqm5}
	\end{equation}
	
	The laser scanning process is modeled by moving the Gaussian distributed heat source on the powder-bed surface at a speed of $\mathbf{v}$. The kinematic equation of temperature is formulated as
	\begin{equation}
	\frac{\partial e}{\partial T} \left[\dot{T}(\mathbf{r},t)-\mathbf{v} \cdot \nabla T \right] + \frac{\partial e}{\partial \rho} \dot{\rho}(\mathbf{r},t) + \sum_\alpha \frac{\partial e}{\partial \eta_\alpha} \dot{\eta}_\alpha(\mathbf{r},t) = \nabla\cdot \left(\mathbf{k}\cdot\nabla T\right)+q(\mathbf{r}),
	\label{eqm6}
	\end{equation}
	where the thermal conductivity $\mathbf{k}$ adopts spatial distribution as $\mathbf{M}$ in Eq. (\ref{eqmM}) i.e. 
	\begin{equation}
	\mathbf{k}=\mathbf{k}_\text{bk}+\mathbf{k}_\text{at}+\mathbf{k}_\text{sf}+\mathbf{k}_\text{gb}.
	\label{eq24}
	\end{equation}
	$q(\mathbf{r})$ represents the volumetric energy deposition due to the radiative energy flux of the laser and can be formulated as \cite{gusarov2005modelling,Gusarov2009}
	\begin{equation}
	q(\mathbf{r})=-\beta P_0p_{xy}(x,y)p_z(\alpha,\lambda,z),
	\label{eqm7}
	\end{equation}
	in which $P_0$ is the nominal laser power reaching the surface of powder bed, $p_{xy}(x,y)$ is the 2D Gaussian distribution. $p_z(\alpha,\lambda,z)$ is a penetration function, which takes the form proposed by Gusarov \textit{et al.} \citep{Gusarov2009}. 
	The hemispherical reflectivity $\alpha$ indicates the influence of the powder material. The attenuation coefficient $\beta$ and the optical thickness $\lambda$ reflect the influence from the powder bed structure. For a loosely packed powder bed with particles of distributed diameters $d$, a thickness of $H_\text{pb}$, and a porosity of $\varepsilon$, the effective value of these two parameters are calculated as
	\begin{equation}  
	\langle\beta\rangle=\frac{3}{2}\frac{1-\varepsilon}{\varepsilon}\langle\frac{1}{d}\rangle, ~ \langle\lambda\rangle=\frac{3}{2}\frac{1-\varepsilon}{\varepsilon}\langle\frac{H_\text{pb}}{d}\rangle.
	\label{eqm8}
	\end{equation}
	
	Since the details inside the substrate is trivial in this work. we set a heat conduction boundary condition (BC) to replace the mesh of the a substrate (Fig. \ref{fig1}a), which has a similar fashion to the third-type BC of heat transfer problems, i.e.
	\begin{equation}
	-(\mathbf{k}\cdot\nabla T) \cdot\mathbf{n}=-\frac{k_\text{sub}}{H_\text{sb}}(\left.T\right|_{\Gamma_\text{sub} }-T_\text{P}),
	\label{eqbc3}
	\end{equation}
	where $\mathbf{n}$ is the normal vector of the boundary $\Gamma_\text{sub}$. $H_\text{sb}$ and $H_\text{sb}$ are the thickness and the heat conductivity constant of the homogeneous substrate, respectively. The bottom of the substrate is set with a fixed temperature $T_\text{P}$ as the pre-heating temperature. 
	
	\subsection{Optimized profile of order parameters}
	It is anticipated that a general powder bed would have a spatial and temporal conserved OP $\rho(\mathbf{r},t)$ in combination with the multiple non-conserved OPs $\{\eta_\alpha\}=\{\eta_1(\mathbf{r},t),\eta_2(\mathbf{r},t),\dots \eta_N(\mathbf{r},t)\}$. The number of non-conserved OPs $N$ is required to be the same as the number of the particles/grains to statistically retain the uniqueness of each crystalline orientation \citep{FAN1997611,KIM2006}. However, such computation is expensive due to the large amount of variables, and inefficient since only a few of variables have nonzero value at any point in the domain. To solve those problems, methods that can drastically reduce the computational cost while retaining the uniqueness of the particles/grains have been proposed in recent decade \citep{KRILLIII20023059,Vedantam2006,permann2016order}. The fundamental idea is to assign the same OP to grains which are sufficiently spaced, and remap the OPs when certain grains with an identical OP tend to have coalesced. This idea makes it possible to use less amount of non-conserved OPs to simulate the evolution of the polycrystalline structure. Taking the example of the grain tracking algorithm proposed by Permann \textit{et al.} in \cite{permann2016order}, it is able to simulate more than 1,000 grains, and only a minimum of 8 non-conserved OPs in 2D and 28 in 3D are required for the grain growth problems. Due to features of SLS, like on-site heating and rapid cooling, global and long-term grain coalescence is absent, in contrast to the grain growth process. This means usually no severe changes of the profile of non-conserved OPs take place, making it possible to further reduce the computational cost by optimizing the profile of OPs.
	
	In this work, we translate the problem of OPs' assignment into a classical minimum coloring problem (MCP). By solving the MCP with the Welsh-Powell algorithm \cite{welsh1967upper}, we obtain the optimized profile with a minimum number of non-conserved OPs. The work flow of the process is presented in Fig. S1a (Supplementary Information). In the beginning, an adjacency matrix of the powder bed (Fig. \ref{fig1}c), demonstrating the ``adjacent neighbors'' of each powder, is generated according to the criterion of adjacency. If two particles/grains which are spaced less than this critical distance, they can be considered as ``adjacent neighbors''. Here the distance between the largest powder and its next-nearest neighbor in hexagonal close packing (inset of in Fig. \ref{fig1}c) is considered as the criterion of adjacency. Then, the map $G(\mathbb{V},\mathbb{E})$ is established based on the adjacency matrix, which consists the unique particle $\nu_i(\mathbf{r},R)$ as the vertices $\mathbb{V}\{\nu_i\}$ and corresponding degree of adjacency $\mathbb{E}\{\epsilon_i\}$, i.e. there are $\epsilon_i$ particles are adjacent to the one indexed as $\nu_i$ (Fig. \ref{fig1}d). Once the map is established, the algorithm will iteratively assign the available non-conserved OPs to non-adjacent and non-assigned particles till all of them have been assigned with OPs. The particles indices remain identical during this iterative assignment, and can be later translated by the grain tracking algorithm (Fig. \ref{fig1}e and \ref{fig1}f). In this work, for the powder bed with particle diameters ranging from 20 to 50 \si{\micro\meter}, 8 non-conserved OPs are sufficient to represent about 200 particles/grains. For another simulation attempt where a loosely packed powder bed with about 400 uniformly sized particles is utilized, only 6 non-conserved OPs are sufficient. In this way, the number of non-conserved OPs is remarkably reduced, and the computation resource is efficiently saved.

	
	\section{Results}\label{Results}
	
	\begin{figure*}[!b]
		\centering
		\includegraphics[width=18cm]{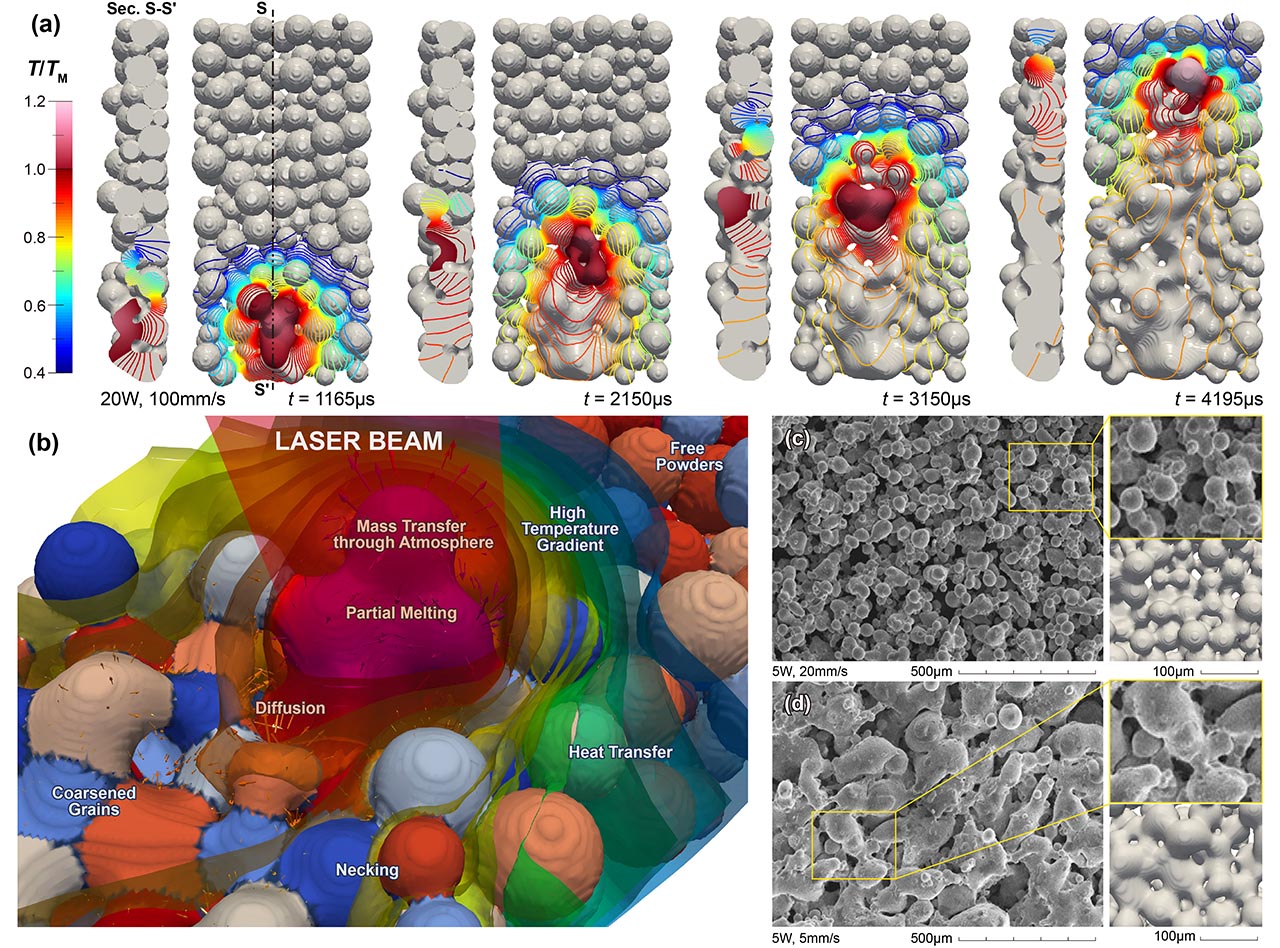}
		\caption{(a) Simulation results on SLS processing of 316L SS powder bed with a laser power of 20 W and scanning speed of 100 mm/s. Region with $T/T_\text{M}>1$ is noted as continuous colormap while one with $T/T_\text{M}<1$ is noted as isotherms; (b) captured phenomena and characteristics around the laser beam spot during the process; Comparison of the surface morphology between the simulated and experimental results in the case of (c) laser power 5 W, scanning speed 20 mm/s and (d) laser power 5 W, scanning speed 5 mm/s. SEM images are reprinted with permission from F.R. Liu \textit{et al.}, J. Mater. Process. Tech., 212 (2012).}
		\label{fig2}
	\end{figure*}
	
	Here we present the simulation results on the microstructure and surface morphology evolution by using the experimental data of the type 316L stainless steel (SS316L) in the inert atmosphere. 
	Assuming the isotropic diffusion and grain boundary migration, the mobilities $\mathbf{L}$ and $\mathbf{M}$ are formulated as scalars and adopted from the self-diffusivity $D_\text{path}^{\text{eff}}$ through possible paths (path $=$ bk, vp, sf and gb) and grain boundary mobility $G_\text{gb}^{\text{eff}}$ \cite{moelans2008quantitative,yang2018arxiv,millett2012phase}, i.e.
	\begin{equation}
	M_\text{path}^{\text{eff}}=\frac{{D}_\text{path}^{\text{eff}}}{2(\underline{C}+\underline{D})},~ L^{\text{eff}}=\frac{G_\text{gb}^{\text{eff}}\gamma_\text{gb}}{T\kappa_\eta}.
	\end{equation} 
	Details of the material-related quantities are elaborated in Note 1 (Supplementary Information), and the procedure to obtain model parameters ($\underline{A},\underline{B},\underline{C}_\text{pt}, \underline{C}_\text{cf}, \underline{D}_\text{pt}, \underline{D}_\text{cf},~\kappa_\eta$ and $\kappa_\rho$) are elaborated in Note 2 (Supplementary Information). 
	The thickness of the powder bed with SS316L particles and the SS316L substrate is $H_\text{pb}=65~\si{\micro\meter}$ and $H_\text{sb}=1000~\si{\micro\meter}$, respectively. The SLS atmosphere is argon. The powder bed consists of particles with distributed diameters $d$ ranging from $20$ to $50~\si{\micro\meter}$. The distribution of $d$ is shown in Fig. S1b. (Supplementary Information). According to Eq. (\ref{eqm8}), the effective attenuation coefficient and optical thickness of the powder bed are calculated as $\langle\beta\rangle=~0.089$ and $\langle\lambda\rangle=5.8$. The Gaussian distribution of the laser beam $p_{xy}(x,y)$ has the full width at half maximum $\text{FWHM}=100~\si{\micro\meter}$. Note that the surface integral of $p_{xy}(x,y)$ reaches $75.8\%$ of $P_0$ in the region with a diameter of FWHM, and reaches $99.5\%$ in the region with a diameter of $2\times\text{FWHM}$. So we regard the region $2\times\text{FWHM}$ as the nominal beam spot. Temperature of the powder bed is initialized as $T/T_\text{M}=0.4$ with the melting point $T_\text{M}=1700~\si{K}$. 
	
	\begin{figure*}[!t]
		\centering
		\includegraphics[width=16cm]{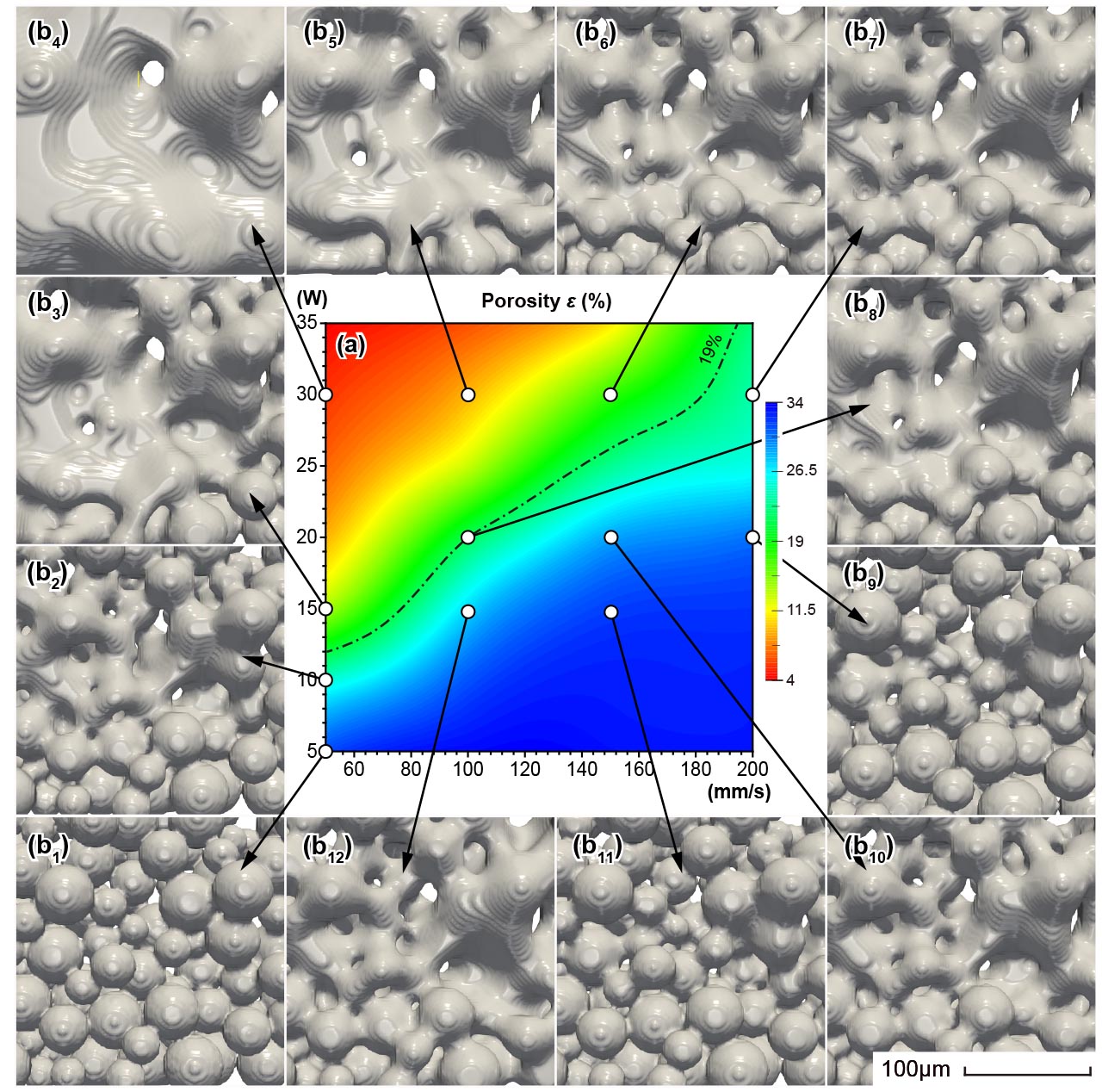}
		\caption{Porosity map (a) and morphology of the SLS processed powder bed (b$_1$)-(b$_{12}$) using different laser power and scanning speed combinations. The dot-dash line is the isoline of 19\% porosity, when 50\% of the initial porosity is reduced.}
		\label{fig3}
	\end{figure*}
	
	Fig. \ref{fig2}a presents the microstructure evolution obtained at a laser power of 20 W and a scanning speed of 100 \si{mm/s}. For realizing the SLS process, a laser beam then scans the powder layer straightly and binds the free particles together simultaneously to form a continuous but porous morphology. The SLS induced physical phenomena around the laser beam spot are shown in Fig. \ref{fig2}b. Due to the laser scanning, temperature of the powder bed obviously increases, as shown in Fig. \ref{fig2}a. Some regions even have the temperature over the melting point $T_\text{M}$, creating the thermodynamic condition for the physical processes such as diffusion and partial melting. The temperature is inhomogeneously distributed and only the particles around the laser beam spot are partially melted. In the region with $T>T_\text{M}$, the reduction of surface energy (or surface capillary pressure) makes the melts flow from the convex to the concave region. This kind of melt flow results in a continuous region with coarsened grains once cooled down to $T<T_\text{M}$. In the region with a maximum temperature lower than $T_\text{M}$, no melting occurs. But the temperature is still high enough to activate diffusion and thus induce necking among adjacent particles. In addition, from the isotherms in Fig. \ref{fig2}a, we can estimate the local temperature gradient around the partial melting region and the pore region as high as $50~\si{K/\micro\meter}$ and $100~\si{K/\micro\meter}$, respectively. Such large-gradient temperature field is essential to the non-isothermal behaviors of the SLS, which could influence the grain coalescence and eventually affect the microstructure evolution \citep{yang2018arxiv}.

	Laser power and scanning speed are important SLS parameters through which one can effectively control the properties (e.g. porosity and grain size) of the processed components. Under a constant laser power of 5 W, we present the simulated surface morphologies along with the corresponding experimental observations in Fig. \ref{fig2}c and \ref{fig2}d for a scanning speed of 20 and 5 mm/s, respectively. It is obvious that changing the scanning speed has significant effects on the attained morphology. For the scanning speed of 20 mm/s in Fig. \ref{fig2}c, powder bed shows a poor binding and the necking among adjacent particles is very weak. For the scanning speed of 5 mm/s in Fig. \ref{fig2}d, on the other hand, a continuous piece form after the SLS processing, indicating the better binding of particles.

	Furthermore, we investigate the influence of scanning speed and laser power by performing a series of simulations. As shown in Fig. \ref{fig3}, the morphology and porosity of the SLS processed powder bed are presented with different laser power and scanning speed.  
	The porosity map can be roughly divided into two regions by the dot-dash line as shown in Fig. \ref{fig3}a. The lower-right region shows less bound particles, while the upper-left region shows more continuous region (or even fully melted as shown in Fig.\ref{fig3}b$_4$). When decreasing the scanning speed while fixing the laser power, more particles are bound to create more continuous pieces, i.e. less porosity will be achieved in the final components (comparing Fig. \ref{fig3}b$_4$-\ref{fig3}b$_7$; \ref{fig3}b$_8$-\ref{fig3}b$_{10}$; and \ref{fig3}b$_3$, \ref{fig3}b$_{12}$-\ref{fig3}b$_{11}$). It is similar for the case when scanning speed is fixed and the laser power is 
	increased (comparing Fig. \ref{fig3}b$_1$-\ref{fig3}b$_4$; \ref{fig3}b$_5$, \ref{fig3}b$_8$, \ref{fig3}b$_{12}$; and \ref{fig3}b$_6$, \ref{fig3}b$_{10}$-\ref{fig3}b$_{11}$). 
	
	In the next section, we would explicitly discuss how laser power and scanning speed influence other features of the powder bed during SLS, including the temperature profile, particles/grains geometry evolution, and microstructure densification.
	
	
	\section{Discussion}\label{Discussion}
	\subsection{Temperature profile}
	
	\begin{figure*}[!t]
		\centering
		\includegraphics[width=18cm]{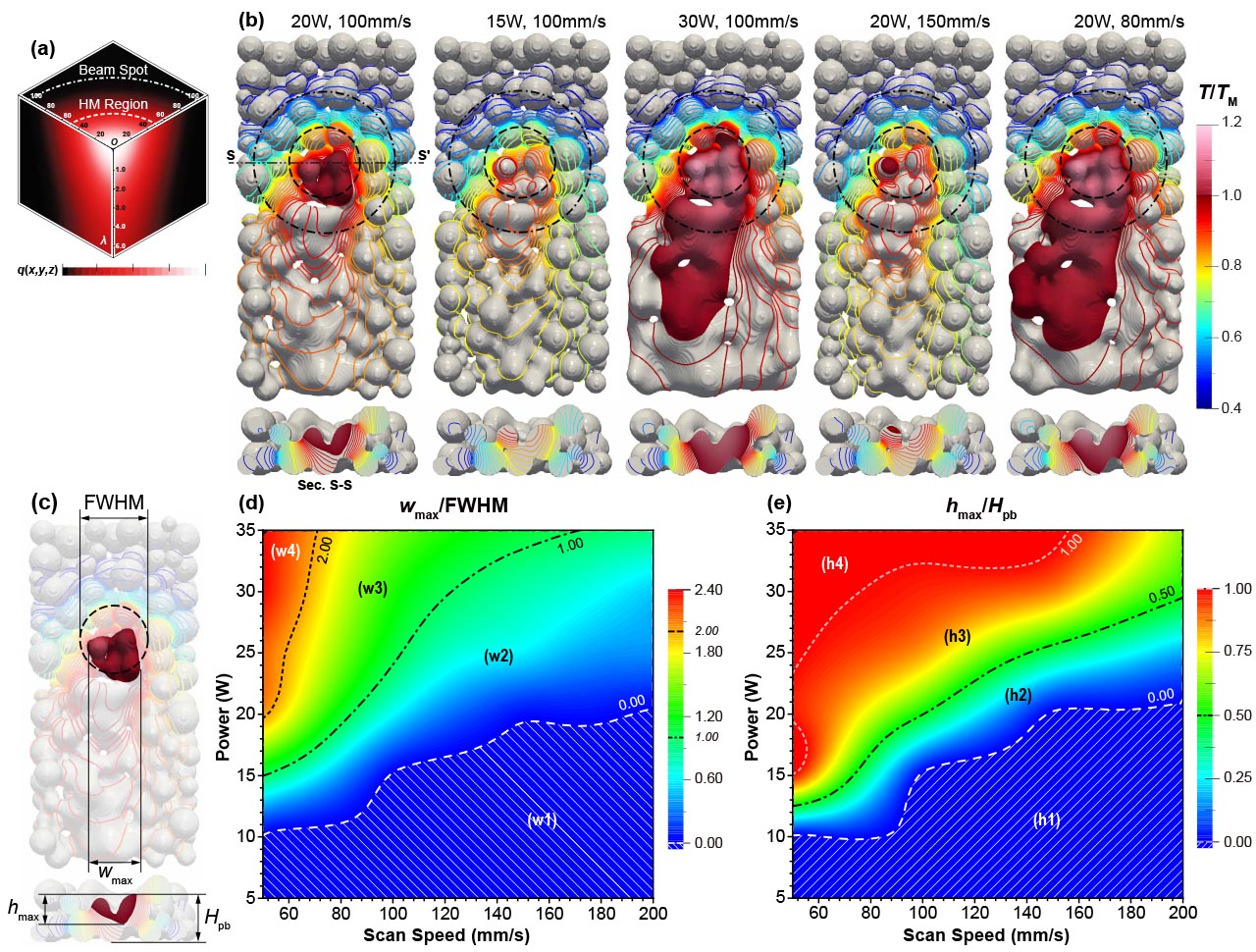}
		\caption{(a) Schematics for distribution of the volumetric energy deposition. (b) Temperature field at the point when the laser been has passed 350 \si{\micro\meter} on the powder bed with different processing parameters. Region with $T/T_\text{M}>1$ is noted as continuous colormap while one with $T/T_\text{M}<1$ is noted as isotherms.  (c) Geometry of partial melting region with width $w$ and depth $h$. (d) Maps of the normalized melting width and (e) normalized melting depth with different combinations of the processing parameters. }
		\label{fig4}
	\end{figure*}
	
	Temperature profile in SLS is directly related to the processing parameters. It is an important indicator for judging whether certain physical process (e.g. partial melting) could occur during SLS. However, due to the difficulties in modeling the powder bed topography and the underlying physics strongly coupled with the heat transfer, simplifications are often used in simulating the temperature profile during powder bed based AM. For instance, the powder bed is frequently assumed to be a homogeneous bulk material with effective thermal properties \cite{dong2009three,Gusarov2009,arisoy2019modeling,gusarov2003contact}. There is also study in the particle scale to simulate the thermal evolution in a 3D multi-layer powder stacking model \cite{liu2012micro}, but the `particles' are kept in the cubic shape and packed regularly. In our work here, the powder bed generated by DEM consists of randomly packed particles. Moreover, the substance field $\rho$ is introduced to assign different thermal properties to the substance and pores separately. The microstructure descriptor $\rho$ and $\{\eta_{\alpha}\}$ are also fully coupled with temperature. In this way, we can not only model the temperature field in the particle scale in a way close to the realistic setup, but also predict the thermally coupled microstructure evolution.
	
	Fig. \ref{fig4}a shows the profile of volumetric energy deposition from the laser beam, in which the half-maximum (HM) region is indicated by the dash line (diameter of FWHM), and the nominal beam spot by the dash-dot line (diameter of $2\times\text{FWHM}$). Fig. \ref{fig4}b presents the top view and the section view of the temperature profile at the point when the laser beam has passed 350 \si{\micro\meter} on the powder bed with different processing parameters. It generally shows an inhomogeneous distribution of isotherms on the powder bed. In the front of the moving beam spot, isotherms are very dense and become sparse once the beam spot has passed away, indicating rapid heating up and a relatively slow cooling down. In this regard, increasing the laser power and decreasing the scanning speed can both improve the heat accumulation around the beam spot. In the case of $P_0=20~\si{\watt}$ and $v=100~\si{\milli\meter/s}$, we can only see some particles are partially melted. When increasing the laser power (e.g. $P_0=30~\si{\watt}$ and $v=100~\si{\milli\meter/s}$ in Fig. \ref{fig4}b) or decreasing the scanning speed (e.g. $P_0=20~\si{\watt}$ and $v=80~\si{\milli\meter/s}$ in Fig. \ref{fig4}b), the melting becomes more noticeable and the partial melting is even extended to full melting. In these cases mechanisms of melting and solidification dominate. On the other hand, decreasing the laser power (e.g. $P_0=15~\si{\watt}$ and $v=100~\si{\milli\meter/s}$ in Fig. \ref{fig4}b) or increasing the scanning speed (e.g. $P_0=20~\si{\watt}$ and $v=150~\si{\milli\meter/s}$ in Fig. \ref{fig4}b) can effectively reduce the melting, and in these cases the mechanism of solid-state sintering is dominant.

	To systematically analyze the influence of processing parameters on the temperature profile and classify the possible dominant mechanism involved in SLS processing, we map the width (normalized with FWHM) and the depth (normalized with $H_\text{pb}$) of the melt with processing parameters, as shown in Fig. \ref{fig4}d and \ref{fig4}e. To help the discussion, each map is divided into four regions and the processing parameters inside a certain region are corresponding to similar microstructure features. For the normalized melting map (Fig. \ref{fig4}d) we have: (w1) no melting; (w2) extremely localized melting within the HM region (diameter of FWHM); (h3) wider melting region, yet still localized within the beam spot (diameter of $2\times\text{FWHM}$); (w4) large-area melting. For the normalized melting depth map (Fig. \ref{fig4}e) we have: (h1) no melting; (h2) partial melting of the particles; (h3) some particles are fully melted and the melt depth exceeds half of the powder bed thickness; (h4) the powder bed is fully melted. Therefore, there is only dominant solid-state sintering in the overlapped region of (w1) and (h1). The dominant melting and solidification occur in regions (w4) and (h4). In the regions (w2), (w3) and (h2) and (h3), particles coexist with melts and thus the liquid-state sintering should be dominant. However, in the overlapped region of (w2) and (h2), the partial melting is restricted to the individual particle and then quickly regrows into the grain after cooling down. This process is closer to solid-state sintering. In other regions, there are more melts mixed with the unmelted particles, leading to typical liquid-state sintering.
	
	\subsection{Particles/grains geometry evolution}
	
	Due to the complex temperature profile, particles/grains during SLS may undergo different physical processes depending on the local thermodynamic conditions. Some particles around the beam spot may suffer from the partial melting. Others may still be sintered together since the temperature is sufficiently high to activate the diffusion and the grain boundary migration. Sorts of physics eventually lead to the evolution in particles/grains geometry evolution, including the change of grain size and grain shape. Such evolution, however, differs hugely from that during the conventional sintering where the temperature is firmly controlled. Since the high temperature gradient exists (there is in maximum $50 ~\si{K/\micro\meter}$ within the particles and $100 ~\si{K/\micro\meter}$ within the pores around the sintered neck), the diffusivity (or grain boundary mobility) would have a very large difference between the hot and the cool ends of particles around the partial melting region. For instance, the surface diffusivity at the hottest end of a partial melted particle is about 100 times larger than on at the coolest end in this work. According to our latest investigation in \cite{yang2018arxiv}, highly non-isothermal condition can also lead to the difference in thermodynamic stability of grains. With the help of the grain tracking algorithm, we can track the particle/grain located at any position in the powder bed and investigate the geometric evolution during the process.
	
	\begin{figure*}[!t]
		\centering
		\includegraphics[width=18cm]{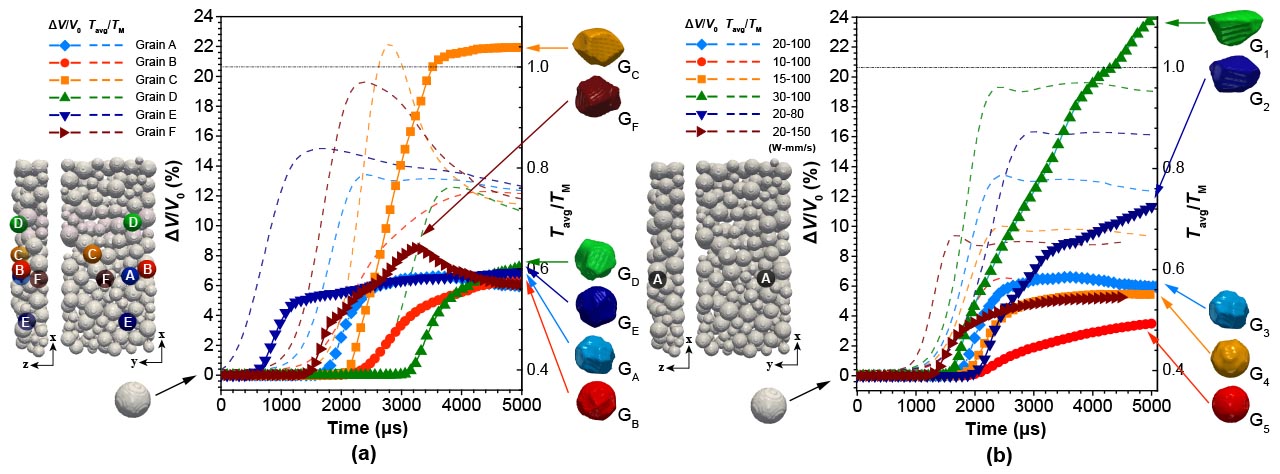}
		\caption{(a) The normalized size variation of the particles/grains located at positions A-F as shown in left inset under the processing parameters of $P_0=20~\si{\watt}$, $v=100~\si{\milli\meter/s}$. Final geometries of corresponding particles/grains are presented in insets g$_\text{A}$-g$_\text{F}$. (b) The normalized size variation of the particles/grains located at position A as shown in left inset under various processing parameters. Final geometries of selected particles/grains are presented in g$_1$-g$_5$.}
		\label{fig5}
	\end{figure*}
	
	Fig. \ref{fig5}a present the geometry evolution of particles/grains of the same processing parameters of $P_0=20~\si{\watt}$ and $v=100~\si{\milli\meter/s}$. Here the volume and the average temperature of the $i\text{-th}$ particle/grain are calculated as
	\begin{equation}
	V=\int \rho\delta(\Omega_i)\text{d}\Omega,~ T_\text{avg}=\frac{\int \rho T\delta(\Omega_i)\text{d}\Omega}{\int \rho\delta(\Omega_i)\text{d}\Omega},
	\label{eqr1}
	\end{equation} 
	where $\delta(\Omega_i)$ is the delta function which takes unity inside the domain $\Omega_i$ of the $i\text{-th}$ particle/grain. $\Omega$ represents the simulation domain. The volume variation $\Delta V$ is thereby calculated as the absolute difference between the current volume $V$ and the initial volume $V_0$. The volume variation is normalized with respect to their initial values. When the laser beam passes, the average temperatures rise to the peak, then gradually drop. The volume variation, however, starts to rise when the average temperature reaches $0.6T_\text{M}$. After reaching the maximum $T_\text{avg}$, the rate of volume variation gradually drops, and the grain volume tend to be constant. Grains A, D and E (Fig. \ref{fig5}a) with the same $y$ coordinate shows almost the same volume variation. At the end of the scanning (5000 \si{\micro\second}), they are in a polyhedral shape with arched surfaces (g$_\text{A}$ ,g$_\text{D}$ and g$_\text{E}$), presenting the resemblance to the grains during the intermediate stage of the sintering. 
	Grain A, B and C with the same $z$ coordinate present different geometric. Among these three grains, grain B is located farthest from the melting region and shows smaller volume variation and rate. It eventually turns into a shape which is less polyhedral, but similar to the one formed by pure necking without subsequent packing (g$_\text{B}$). 
	Grain C is located closest to the center of the partial melting region, whose $T_\text{avg}$ firstly rises above $T_\text{M}$ then quickly drops. It eventually turns into a flat grain (g$_\text{C}$) with a smooth top surface (due to partial melting) and polyhedral bottom surfaces (due to packing with other grains) . 
	Grain F is exactly located beneath the partial melting region. Its volume variation increases firstly and drops later. Its volume variation reaches the maximum, then drops after reaching maximum $T_\text{avg}$. In the end, it also turns into a polyhedral shape (g$_\text{F}$).
	
	In Fig. \ref{fig5}b we also present the volume variation and variation rate of the same particle/grain (grain A) with different processing parameters. It can be found that increasing the laser power and decreasing the scanning speed can increase the volume variation. For $v=100~\si{\milli\meter/s}$, the grain is polyhedral at $P_0=20~\si{\watt}$ (g$_3$), but is round with a top-surface necking (g$_5$) at $P_0=10~\si{\watt}$. For the cases of $P_0=30~\si{\watt}$, $v=100~\si{\milli\meter/s}$ and $P_0=20~\si{\watt}$, $v=80~\si{\milli\meter/s}$, flat grains with a smooth top and a polyhedral bottom are obtained (g$_1$ and g$_2$). We can also observe the fluctuation of the volume variation in the case of $P_0=30~\si{\watt}$, $v=100~\si{\milli\meter/s}$ when $T_\text{avg}$ is approaching its maximum, which is close to the melting point. This may be due to the contribution from the partial melting term $\Phi(\tau)\mathbf{M}_\text{melt}$.
	
	\subsection{Microstructure densification}
	
	\begin{figure*}[!t]
		\centering
		\includegraphics[width=18cm]{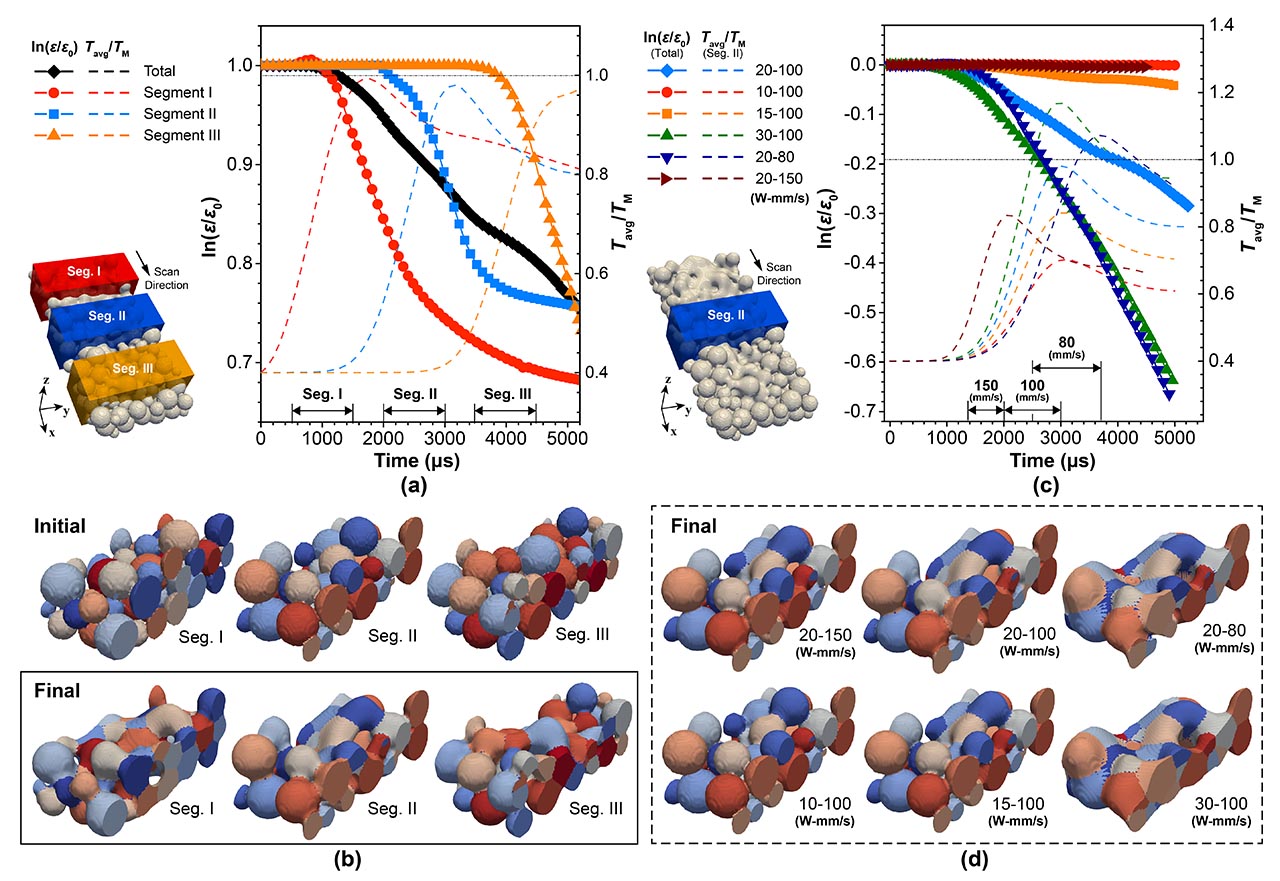}
		\caption{(a) Temporal evolution of $\ln({\varepsilon/\varepsilon_0})$ and the average temperature $T_\text{avg}$ of the different segments and the whole powder bed under the processing parameters of $P_0=20~\si{\watt}$, $v=100~\si{\milli\meter/s}$, and (b) the final microstructures of selected segments. (c) The temporal evolution of $\ln({\varepsilon/\varepsilon_0})$ of the whole powder bed and the $T_\text{avg}$ of the same segment of the powder bed under the various processing parameters, and (d) the final microstructures of the segment. Time windows shown in (a) and (b) indicate the periods when the beam scans on corresponding segment.}
		\label{fig6}
	\end{figure*}
	
	Particle/grain geometric evolution where the grains get coarser and more packed while the pores shrink and reshapes, leads to the microstructure densification, which can be characterized by the porosity evolution. In Ref. \citep{simchi2006direct} Simchi proposed that the temporal evolution of the porosity during the direct laser sintering of metal follows the first order kinetics law as
	\begin{equation}
	\dot{\varepsilon}=-k\varepsilon,
	\label{eqr2}
	\end{equation} 
	where $k$ is defined as the sintering rate constant. In Fig. \ref{fig6}a we present the temporal evolution of $\ln({\varepsilon/\varepsilon_0})$ of the three selected segments and the whole powder bed with an initial porosity $\varepsilon_0$ under the processing parameters of $P_0=20~\si{\watt}$, $v=100~\si{\milli\meter/s}$.
	The average temperature $T_\text{avg}$ of each segment is also calculated according to Eq. (\ref{eqr1}). 
	These three segments are with the same geometry (length 100 \si{\micro\meter}, width 250 \si{\micro\meter}), but located at different positions in the powder bed. It can be seen that when the laser beam scans into a certain segment, $T_\text{avg}$  rises and $\ln({\varepsilon/\varepsilon_0})$ is almost linearly decreased there. Once the laser beam is about to leave the certain segment, the $T_\text{avg}$ reaches its maximum then start to drop, while $\ln({\varepsilon/\varepsilon_0})$ is slowly decreased. 
	In contrast, $\ln({\varepsilon/\varepsilon_0})$ of the whole powder bed presents an approximately linear trend, with a fitted slope as $k=7\times10^{-5}~\si{\second^{-1}}$, as shown in Fig. \ref{fig6}a. This demonstrates that Eq. (\ref{eqr1}) is valid in this model to describe the porosity evolution of the powder bed combining all contributions from each finite segments, which is also in agreement with the experimental measurement in Ref. \citep{simchi2006direct}. According to Fig. \ref{fig6}b, we can see the final microstructure ($t=5000~\si{\micro\second}$) around the segment center shows the resemblance to the microstructure during the intermediate stage of sintering \cite{german2014sintering}. In this stage, grains have been already packed together while most of the open pores have been eliminated or rearranged into a roundly closed shape. Whereas the microstructure around the segment margin presents less packed but more necked grains surrounded by tunnel-like open pores, manifesting the characteristics during the initial stage of sintering \cite{german2014sintering}.
	
	In Fig. \ref{fig6}c we present the temporal evolution of $\ln({\varepsilon/\varepsilon_0})$ of the whole powder bed under different processing parameters. Segment II is selected to present the influence of processing parameters on temporal $T_\text{avg}$, which shows a similar trend as in Fig. \ref{fig6}a. The approximately linear trend of $\ln({\varepsilon/\varepsilon_0})$ is still obvious. Both increasing the laser power and decreasing the scanning speed result in the increase of maximum $T_\text{avg}$ and the rate constant $k$. Final microstructures of the segment II under various processing parameters are shown in Fig. \ref{fig6}d. We set the microstructure of segment II of $P_0=20~\si{\watt}$, $v=100~\si{\milli\meter/s}$
	as the reference and compare it with cases with other processing parameters. For the cases with fixed $v=100~\si{\milli\meter/s}$ and $P_0=10$, $15$ and $20~\si{\watt}$, densification rarely occurs and only necking with tunnel-like open pores form, also manifesting the characteristics of the initial stage of sintering.
	For the case of $P_0=30~\si{\watt}$ and $v=100~\si{\milli\meter/s}$, however, the microstructure is significantly densified with packed grains and almost no pores, showing the characteristics of the final stage of the sintering where the grain coarsening dominates. For the case of $P_0=20~\si{\watt}$ and $v=80~\si{\milli\meter/s}$, the microstructure densification occurs, but there are still some round pores near the segment margin, indicating the transition from the intermediate stage to the final stage of sintering.
	
	\begin{figure*}[!t]
		\centering
		\includegraphics[width=18cm]{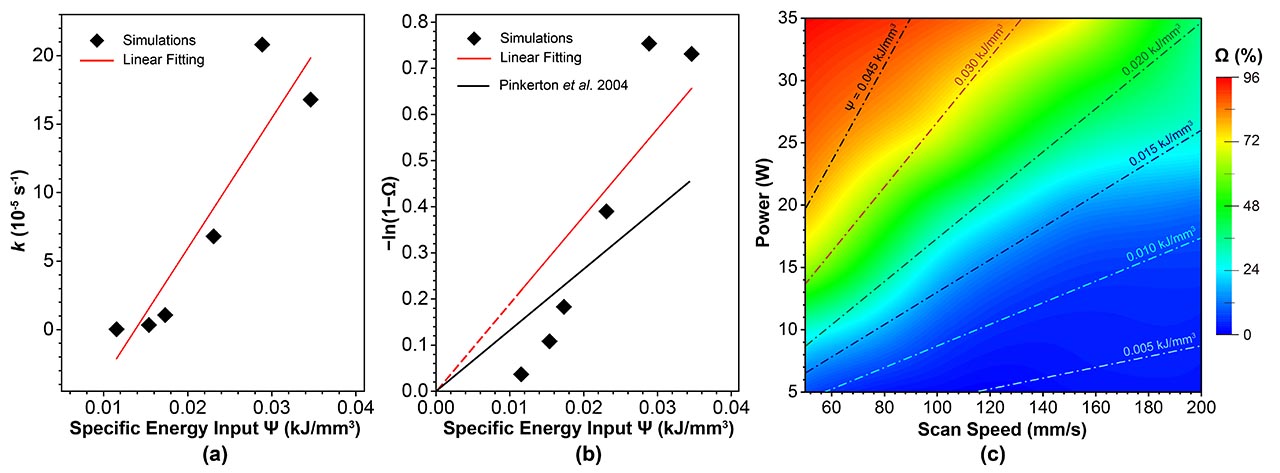}
		\caption{(a) The sintering rate constant $k$ and (b) $-\ln(1-\Omega)$ as a function of the specific energy input $\Psi$. Linear fittings are illustrated by the red lines. (c) Map of densification factor $\Omega$ along with different laser power and scanning speed, where the dot-dash lines denote different specific energy input $\Psi$.}
		\label{fig7}
	\end{figure*}
	
	Here we further discuss the relation between the processing parameters and the sintering rate as well as the final densification. We define the specific energy input $\Psi$ and densification factor $\Omega$ as 
	\begin{equation}
	\Psi=\frac{P}{hwv},~\Omega=\frac{\varepsilon_0-\varepsilon}{\varepsilon_0-\varepsilon_\text{min}}.
	\label{eqr2}
	\end{equation} 
	In Eq. (\ref{eqr2}), $P$ and $v$ is the laser power and the scanning speed of the laser beam, respectively. $h$ and $w$ is the thickness and width of the scan track, respectively. $\varepsilon_\text{min}$ is the minimum attainable porosity in a sintered part, which ranges between 0.02 and 0.3 depending on the material properties. To avoid the deviation due to the less densified margin of the powder bed, we choose a $\text{FWHM}=100~\si{\micro\meter}$ of the power density of the laser as the track width $w$, and $P=75.8\%P_0$ (the surface integral of the power density reaches $75.8\%$ of $P_0$). $h$ takes $H_\text{pb}=65~\si{\micro\meter}$ and $\varepsilon_\text{min}=3\%$. 
	As shown in Fig. \ref{fig7}a, the relation $k$ \textit{vs} $\Psi$ can be linearly fitted with a slope of $9.51\times10^{-3}$. Similarly, linear fitting of $-\ln(1-\Omega)$ \textit{vs} $\Psi$ give the densification coefficient $K=18.97$, as shown in Fig. \ref{fig7}b. It should be noted that $K$ is related to the material properties and the particle diameter distribution in the powder bed \citep{simchi2006direct}. The $K$ value obtained here is in line with the experimental $K=12.6$ in Refs. \citep{pinkerton2004behaviour} and \citep{simchi2006direct} where the SS316L particles are with diameter $<$ 45 \si{\micro\meter}. The agreement in $K$ shows the applicability of the developed model in simulating the microstructure densification. We also present a detailed map of the densification factor $\Omega$ in Fig. \ref{fig7}c. Following the isoline of the specific energy input $\Psi$ (dot-dash lines in Fig. \ref{fig7}c), increasing laser power/scan speed actually leads to the increase of the densification $\Omega$. And when higher $\Psi$ are kept, the faster increase of $\Omega$ can be found. It reveals another interesting fact that specific energy input may not uniquely identify the densification of the processed component during SLS. Similar pattern has been experimentally observed in Ref. \citep{ prashanth2017energy}.
	
	To sum up, we have developed a 3D non-isothermal phase-field model for investigating the evolution of the microstructure during SLS. Through numerical simulations, we capture interesting phenomena during SLS, such as temperature field with high gradient, mass transfer through partial melting, diffusion and evaporation, particle/grain necking and coarsening which leads to the microstructure densification. We also reveal the influences of the processing parameters (i.e. laser power and scanning speed) on microstructural features, including the porosity, surface morphology, temperature profile, particle/grain geometry, and densification. We further present the feasibility of the first-order kinetics during the porosity evolution, and verify the linkage of the sintering rate constant and the densification factor to the specific energy input. It is expected to further investigate the influence from factors such as the scanning strategies, initial particle size distribution and shape during SLS processing in the near future.
	
	\section{Method}\label{Method}
	
	\subsection{Powder bed generation using discrete element method}
	To generate the initial powder bed, the discrete element method (DEM) simulator YADE is used \cite{vsmilauer2015yade}. The spheres are firstly created as the gaseous loose-packing cluster with no contacts between particles. The distribution of the particle diameter is shown in Fig. S1b (Supplementary Information). Then, a gravitational force is imposed on the particles to make them spread into a rectangular box. The iterations continue until the deviation from force equilibrium on the particles is below a certain threshold, i.e the particles are stationary. This process is shown in Fig. S1b (Supplementary Information). After the powder bed is generated, the center $\mathbf{r}$ and radius $R$ of each unique particle is recorded as a vertex $\nu_i(\mathbf{r},R)$, and added into the vertices list $\mathbb{V}\{\nu_i\}$ for further optimization.
	
	\subsection{Implementation of finite element method}
	The model is numerically implemented within the MOOSE framework by finite element method (FEM) \citep{tonks2012object}. 8-node hexahedron Lagrangian elements are chosen to mesh the geometry. The Cahn-Hilliard equation in Eq. (\ref{eqm4}), which is a 4th order differential equation, is solved by splitting it into two 2nd order differential equations via the introduction of an additional coupling field $\mu$ \citep{elliott1989second,zhao2015isogeometric,balay1997efficient}. Then, the weak forms of split Eq.(\ref{eqm4}), along with the Eqs. (\ref{eqm5}) and (\ref{eqm6}), are discretized following the Galerkin method. To solve those time-dependent partial differential equations, 
	transient solver with preconditioned Jacobian-Free Newton–Krylov (PJFNK) method and backward Euler algorithm are employed. Adaptive meshing and time stepping schemes are used to reduce the computation costs. The constraint of the order parameters is fulfilled using penalty functions. 
	
	\subsection{Parallel CPU computation}
	The large-scale parallel CPU computations for each simulation domain, which has degree-of-freedoms (DOFs) on the order of 10,000,000 for both nonlinear system and auxiliary system, are performed with 150 processors and 2 GByte RAM per processor based on OpenMPI. Each  simulation consumes on the order 10,000 of core hours.
	
	\subsection{Data availability}
	The authors declare that the data supporting the findings of this study are available within the paper and its Supplementary Information files.
	
	\section{Acknowledgement}\label{Method}
	The support from the European Research Council (ERC) under the European Union's Horizon 2020 research and innovation programme (grant agreement No 743116) and the Profile Area From Material to Product Innovation -- PMP (TU Darmstadt) is acknowledged. The authors also greatly appreciate their access to the Lichtenberg High Performance Computer and the technique supports from the HHLR, Technische Universit\"at Darmstadt.
	
	\section{Author Contributions}\label{Method}
	Y.Y. performed the phase-field simulations, data processing, and manuscript writing. O.R. performed the powder bed generation and contributed to the manuscript writing. Y.B. supported numerical calculations in HPC. B.-X.X. supervised the project, analyzed the results, and revised the manuscript. M.Y. contributed by intensive consultation and discussion. All authors reviewed and approved the manuscript.

	\newpage
	

	\bibliography{reference}
	
\end{document}



\title{Supplementary Information \\ \textbf{Three-dimensional non-isothermal phase-field modeling of microstructure evolution during selective laser sintering}}
\author[els]{Yangyiwei Yang}

\author[rvt]{Olav Ragnvaldsen}

\author[els]{Yang Bai}

\author[els]{Min Yi\corref{cor1}}
\ead{yi@mfm.tu-darmstadt.de}

\author[els]{Bai-Xiang Xu\corref{cor1}}
\ead{xu@mfm.tu-darmstadt.de}

\cortext[cor1]{Corresponding author}

\address[els]{Mechanics of Functional Materials Division, Institute of Materials Science, Technische Universit\"at Darmstadt, Darmstadt 64287, Germany}

\address[rvt]{Department of Materials Science and Engineering, Norwegian University of Science and Technology, 7491 Trondheim, Norway}

\maketitle
\renewcommand\theequation{S\arabic{equation}}
\renewcommand\thefigure{S\arabic{figure}}
\renewcommand\thetable{S\arabic{table}}

\begin{figure*}[h]
\centering
\includegraphics[width=16cm]{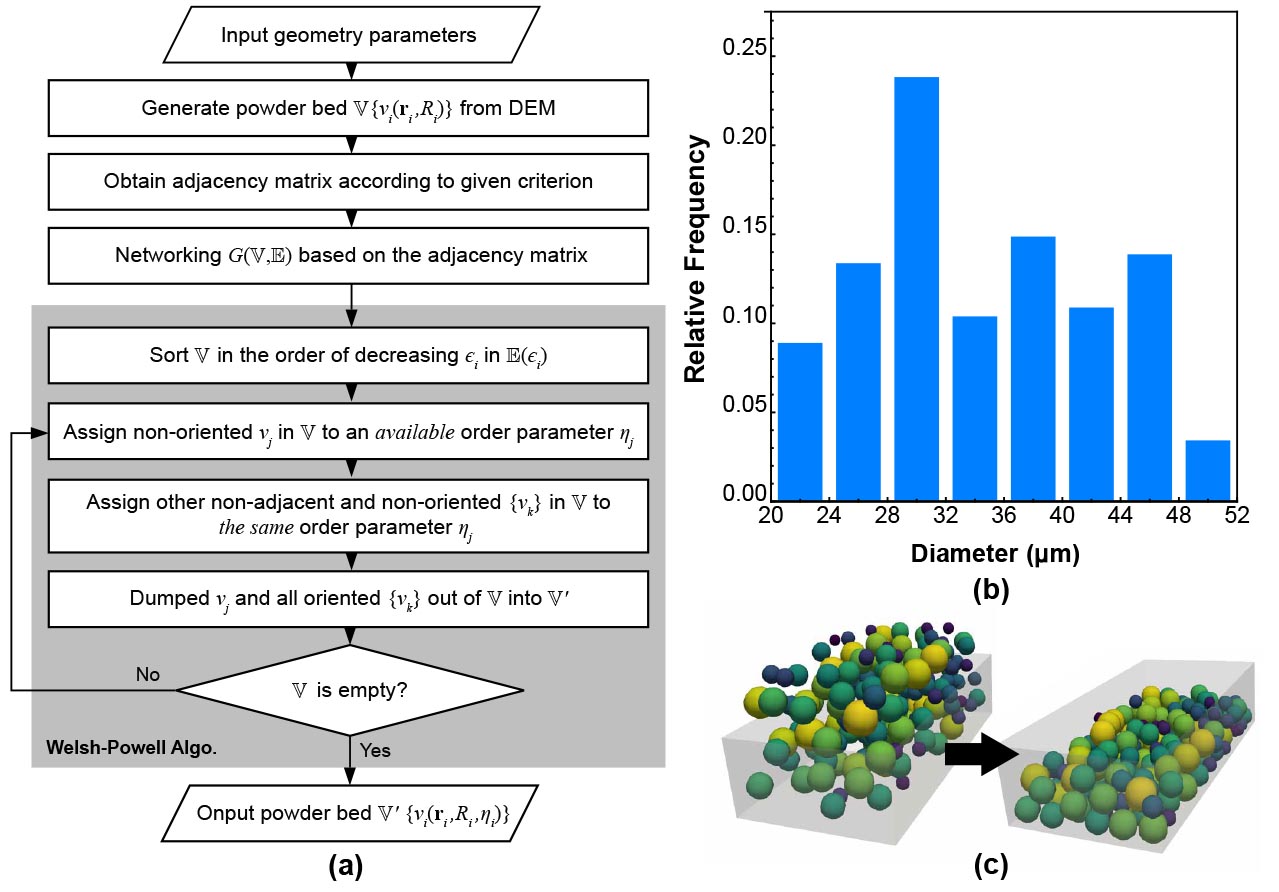}
\caption{(a) Work flow of the optimization steps using the Welsh-Power algorithm; (b) diameter distribution of the powder bed; (c) generation of initial powder bed through gravitational spreading in DEM.}
\label{figs1}
\end{figure*}

\clearpage

\begin{figure*}[!t]
\centering
\includegraphics[width=16cm]{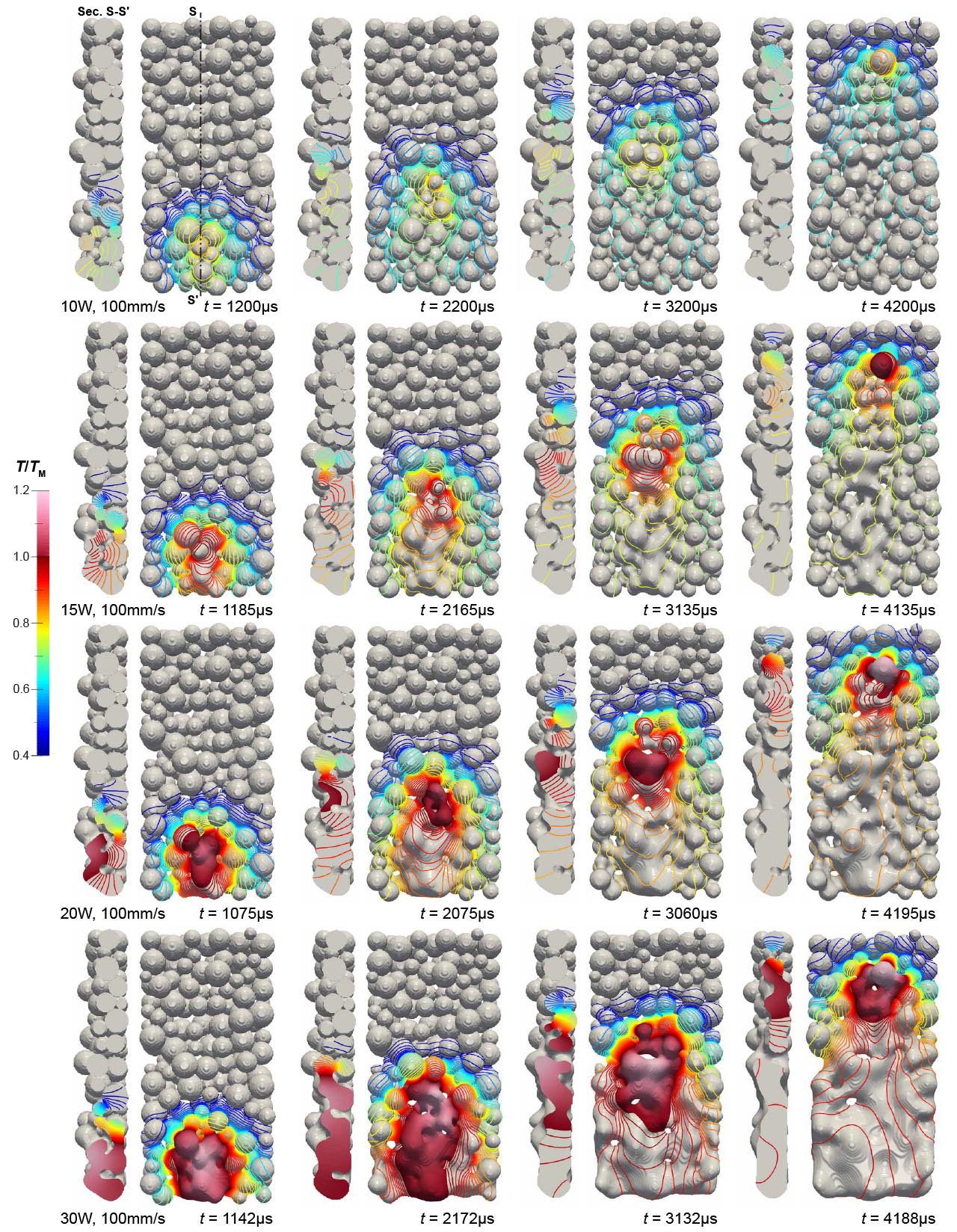}
\caption{Simulation results on SLS processing of the SS316L powder bed with various laser power and a constant scanning speed of 100 mm/s. Region with $T/T_\text{M} > 1$ is noted as continuous colormap while one with $T/T_\text{M} < 1$ is noted as
isotherms.}
\label{figs2}
\end{figure*}

\clearpage

\begin{figure*}[!t]
\centering
\includegraphics[width=16cm]{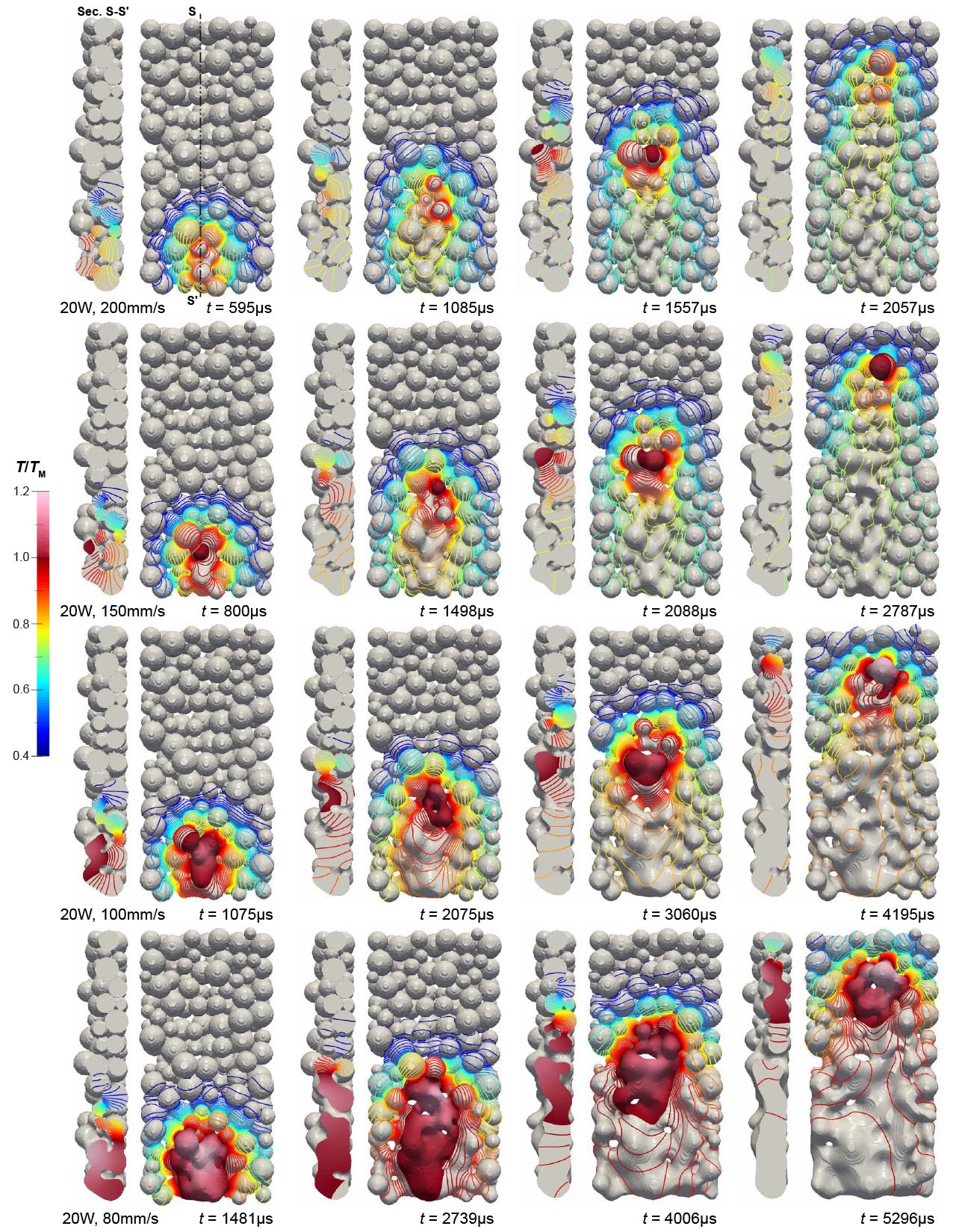}
\caption{Simulation results on SLS processing of the SS316L powder bed with a constant laser power of 20 W and various scanning speeds. Region with $T/T_\text{M} > 1$ is noted as continuous colormap while one with $T/T_\text{M} < 1$ is noted as
isotherms.}
\label{figs3}
\end{figure*}

\clearpage

\subsection*{Supplementary Note 1: Material properties and normalization}
Table \ref{tab1} gives the material properties of SS316L in the argon atmosphere. Table \ref{tab2} gives the dimensionless forms of the quantities utilized in this work. The dimensionless forms are obtained by normalizing with respect to a set of reference quantities, including the reference (melting) temperature $T_\text{M}$, the reference length scale $\lambda$, and the time scale $\tau$. The reference energy density $\underline{C}_\text{pt}^{T_\text{M}}=\kappa_\rho^{T_\text{M}}T_\text{M}/\lambda^2$ which is the model parameter obtained at the reference temperature $T_\text{M}$. $\kappa_\rho^{T_\text{M}}$ is the gradient model parameter at a reference temperature $T_\text{M}$. Other quantities are given in Table \ref{tab2}. Spatial and time derivatives are also normalized with respect to $\tau$ and $\lambda$, respectively. 

\renewcommand{\arraystretch}{1.5}
\begin{table}[h] \small
\centering 
\begin{threeparttable}
\caption{Material properties used in simulations.}
\begin{tabular}{cccccc}
\hline
Properties & Expressions ($T$ in K) & Units &  References \\ \hline
$T_\text{M}$ & $\sim1700$ & K & & \\
$\gamma_\text{sf} 
$ & $10.315-5.00\times 10^{-3}T$ & J/m$^2$ & \citep{price1964surface} \\
$\gamma_\text{gb}
$ & $13.018-7.50\times 10^{-3}T$ & J/m$^2$ & \citep{price1964surface} \\
$D_\text{sf}^\text{eff}$ & $0.40\text{exp}\left(-2.200\times 10^5/\mathfrak{R}T \right)$ & m$^2$/s & \citep{blakely1963studies} \\
$D_\text{gb}^\text{eff}$ & $2.40\times 10^{-3}\text{exp}\left(-1.770\times 10^5/\mathfrak{R}T \right)$ & m$^2$/s & \citep{blakely1963studies} \\
$D_\text{bk}^\text{eff}$ & $2.17\times 10^{-5}\text{exp}\left(-2.717\times 10^5/\mathfrak{R}T \right)$ & m$^2$/s & \citep{mead1956self} \\
$G_\text{gb}^\text{eff}$ & $3.26\times 10^{-3}\text{exp}\left(-1.690\times 10^5/\mathfrak{R}T \right)$ \tnote{*} & m$^4$/(J s) &  \\
$M_\text{melt}^\text{eff}$ & $\sim3.45\times 10^{-13}$ \tnote{**} & m$^5$/(J s) &  \\
$k_\text{316L}^\text{eff}$ & $10.292+0.014T$ & J/(s m K) & \citep{liu2012micro} \\
$k_\text{Ar}^\text{eff}$ & $\sim0.06$ & J/(s m K) & \citep{hoshino1986determination} \\
$c_\text{316L}$ & $3.61\times 10^6+1272T$ & J/(m$^3$ \text{K}) & \citep{liu2012micro} \\
$c_\text{Ar}$ & 717.6 & J/(m$^3$ \text{K}) & \citep{chase1998nist}\\
$\mathscr{L}_\text{316L}$ & $2.4\times 10^9$ & J/$m^3$ & \citep{liu2012micro} \\
$\alpha_\text{316L}$ & 0.7 &   & \citep{Gusarov2009}\\
    \hline
\end{tabular}
\label{tab1}
\begin{tablenotes}
\footnotesize
\item[*] Activation energy is obtained from \citep{di2002analysis} while the prefix factor is estimated as unity at $T_\text{M}$ after normalization.
\item[**] Estimated as $100M_\text{sf}^\text{eff}$.
\end{tablenotes}
\end{threeparttable}
\end{table}

\newcommand{\tabincell}[2]{\begin{tabular}{@{}#1@{}}#2\end{tabular}}
\renewcommand{\arraystretch}{1.5}
\begin{table}[h] \small
\centering 
\caption{The dimensionless form of the quantities involved in this model.}
\begin{tabular}{cccccc}
\hline
 & Symbols & Normalization  &  &  Symbols & Normalization \\ \hline
\tabincell{c}{Physical \\ quantities} & 
\tabincell{c}{  $c$ \\ $k$ \\ $L$ \\ $M$ \\$q$ \\ $v$ \\$\mathscr{L}$} &  
\tabincell{c}{ $\widetilde{c}=cT_\text{M}/\underline{C}_\text{pt}^{T_\text{M}}$ \\ $\widetilde{k}=k \tau T_\text{M} /(\underline{C}_\text{pt}^{T_\text{M}}\lambda^2)$ \\ $\widetilde{L}=L \tau \underline{C}_\text{pt}^{T_\text{M}}$  \\ $\widetilde{M}=M \tau \underline{C}_\text{pt}^{T_\text{M}} / \lambda^2$  \\ 
$\widetilde{q}=q\tau \underline{C}_\text{pt}^{T_\text{M}} / \lambda^3$ \\ 
$\widetilde{v}=v\tau / \lambda$ \\
$\widetilde{\mathscr{L}}=\mathscr{L} /\underline{C}_\text{pt}^{T_\text{M}} $
} & 
\tabincell{c}{Model \\ parameters} & 
\tabincell{c}{$\underline{A}$ \\ $\underline{B}$ \\ $\underline{C}_\text{pt}$ \\ $\underline{D}_\text{pt}$ \\ $\underline{C}_\text{cf}$ \\ $\underline{D}_\text{cf}$ \\ $\kappa_\rho$ \\ $\kappa_\eta$ }  & 
\tabincell{c}{$\widetilde{\underline{A}}=\underline{A}$ \\ $\widetilde{\underline{B}}=\underline{B}$ \\ $\widetilde{\underline{C}}_\text{pt}=\underline{C}_\text{pt} / \underline{C}_\text{pt}^{T_\text{M}}$ \\ $\widetilde{\underline{D}}_\text{pt}=\underline{D}_\text{pt} / \underline{D}_\text{pt}^{T_\text{M}}$ \\ $\widetilde{\underline{C}}_\text{cf}=\underline{C}_\text{cf} T_\text{M} / \underline{C}_\text{pt}^{T_\text{M}}$ \\ $\widetilde{\underline{D}}_\text{cf}=\underline{D}_\text{cf} T_\text{M} / \underline{C}_\text{pt}^{T_\text{M}}$ \\ $\widetilde{\kappa}_\rho=\kappa_\rho T_\text{M} / (\underline{C}_\text{pt}^{T_\text{M}}\lambda^2)$ \\ $\widetilde{\kappa}_\eta=\kappa_\eta T_\text{M} / \underline{C}_\text{pt}^{T_\text{M}}\lambda^2)$} \\
    \hline
\end{tabular}
\label{tab2}
\end{table}

\newpage

\subsection*{Supplementary Note 2: Obtaining model parameters by fitting the temperature-dependent surface and interface energies}
In this section, we present the derivation of the surface and grain-boundary energies at equilibrium which is based on the previous works by \citep{61moelans2008quantitative}, \citep{474kleinlogel2000sintering}, and \citep{35fan1997effect}. Here we firstly show the dependency of the model parameters from Eq. (\ref{eqA1}) to (\ref{eqA8}), then derive the explicit formulation of the surface and grain-boundary energies from Eq. (\ref{eqA9}) to (\ref{eqA17}). Finally, direct and indirect methods of determining eight model parameters are proposed in the rest of this appendix.

\begin{figure}[!t]
    \centering
    \includegraphics[width=16cm]{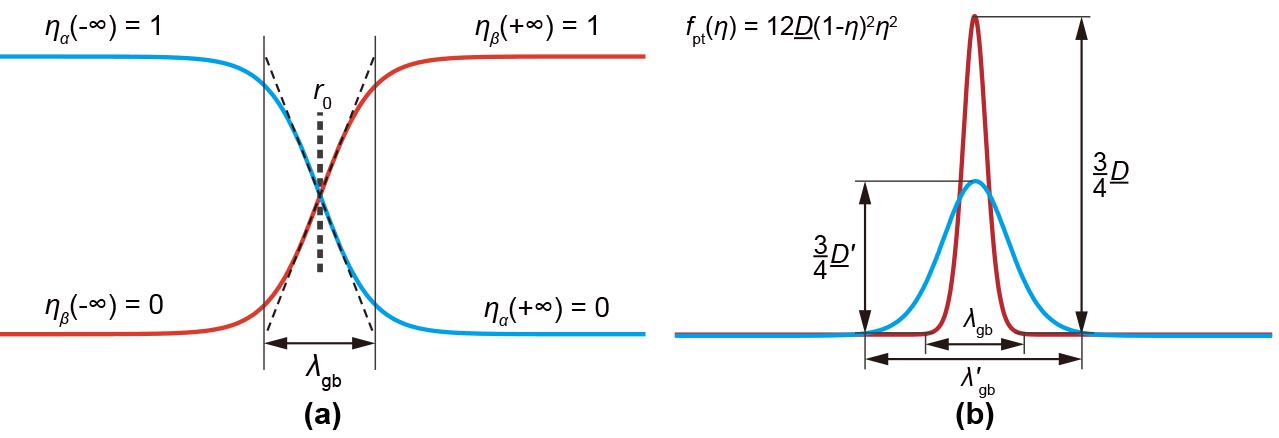}
    \caption{(a) Schematic of the diffusive grain boundary and the profile of $\{\eta_\alpha\}$ across such interface. (b) Schematic of the potential part of the free energy density $f_\text{pt}=12\underline{D}\left(1-\eta\right)^2\eta^2$ with a fixed $\gamma_\text{gb}$ and different ${\lambda }_{\text{gb}}$.}
    \label{figA1}
\end{figure}

Considering the profiles of order parameters across the surface of a grain, where $\rho$ and only one $\eta$ varies from a semi-finite atmosphere/pore phase ($-\infty$) to a semi-finite grain phase ($+\infty$), as shown in Fig. \ref{figA1}. As for the normal direction $\mathbf{r}$ of an arbitrary point on the surface, profiles of $\rho$  and $\eta$ should satisfy the boundary conditions
\begin{equation}  
\left\{  
\begin{array}{lr}  
\rho=\eta=0  &  r \rightarrow -\infty  \\  
\rho=\eta=1  &  r \rightarrow +\infty  \\  
\nabla_\textbf{r}\rho = \nabla_\textbf{r}\eta = 0 & r \rightarrow \pm\infty 
\end{array}  
\right.
\label{eqA1}
\end{equation} 
Then the specific surface energy can be calculated by Cahn’s approach \citep{33cahn1961spinodal,38cahn1958free}
\begin{equation}
\gamma_\text{sf}=\int_{-\infty}^\infty \left[ f(T,\rho,\{\eta,0\})\right]\text{d}\mathbf{r},
\label{eqA2}
\end{equation}
where
\begin{equation*}
\begin{split}
f(T,\rho,\{\eta_\alpha\})=& \xi f_\text{ht}(T)\left(\underline{A}\rho + \underline{B}\sum_\alpha\eta_\alpha\right) +
\underline{C}\left[\rho^2(1-\rho)^2 \right] + \\
&\underline{D}\left[\rho^2+6(1-\rho)\sum_\alpha\eta_\alpha^2 - 4(2-\rho)\sum_\alpha\eta_\alpha^3 + 3\left(\sum_\alpha\eta_\alpha^2 \right)^2 \right] + \\
&\frac{1}{2} T \kappa_\rho \left| \nabla_\mathbf{r} \rho \right|^2 + \frac{1}{2} T \kappa_\eta \sum_\alpha \left| \nabla_\mathbf{r} \eta_\alpha \right|^2,
\label{eqm2}
\end{split}
\end{equation*}
At equilibrium, functional in Eq. (\ref{eqA2}) maintains minimum at each temperature $T$, requiring $\rho$ and $\eta$ to satisfy the Euler-Lagrange equation

\begin{equation}
\frac{\partial f(T,\rho,\{\eta,0\})}{\partial\rho}-T\kappa_\rho\nabla^2_\mathbf{r}\rho = \frac{\partial f(T,\rho,\{\eta,0\})}{\partial\eta}-T\kappa_\eta\nabla^2_\mathbf{r}\eta = 0 .
\label{eqA4}
\end{equation}
In this model, the constraint of order parameters $(1-\rho)+\sum_{\alpha} \eta_\alpha=1$ should hold in any region within the substance at any time. Assuming $\rho$ and $\eta$ adopt the same shape as shown in Fig. \ref{figA1}a, i.e. $\rho(\mathbf{r})=\eta(\mathbf{r})$, from Eq. (\ref{eqA4}) we can yield the following equation
\begin{equation}
\frac{1}{{{\kappa }_{\rho }}}\frac{\partial f\left( T,\rho ,\left\{ \eta ,0 \right\} \right)}{\partial \rho }=\frac{1}{{{\kappa }_{\eta }}}\frac{\partial f\left( T,\rho ,\left\{ \eta ,0 \right\} \right)}{\partial \eta },
\label{eqA5}
\end{equation}
where
\begin{equation*}
    \left\{ \begin{aligned}
        & \frac{\partial f\left( T,\rho ,\left\{ \eta ,0 \right\} \right)}{\partial \rho }=\underline{A}{{f}_{\text{ht}}}\left( T \right)+\underline{C}\left( 2\rho -6{{\rho }^{2}}+4{{\rho }^{3}} \right)+\underline{D}\left( 2\rho -6{{\eta }^{2}}+4{{\eta }^{3}} \right), \\ 
        & \frac{\partial f\left( T,\rho ,\left\{ \eta ,0 \right\} \right)}{\partial \eta }=\underline{B}{{f}_{\text{ht}}}\left( T \right)+\underline{D}\left[ 12\left( 1-\rho  \right)\eta -12\left( 2-\rho  \right){{\eta }^{2}}+12{{\eta }^{3}} \right]. \\ 
    \end{aligned} 
    \right.
\end{equation*}
Replacing every $\eta$ by $\rho$, we obtain
\begin{equation}
\frac{1}{{{\kappa }_{\rho }}}\left[ \underline{A}{{f}_{\text{ht}}}\left( T \right)+\left( \underline{C}+\underline{D} \right)\left( 2\rho -1 \right)\left( 2\rho -2 \right)\rho \right]=\frac{1}{{{\kappa }_{\eta }}}\left[ \underline{B}{{f}_{\text{ht}}}\left( T \right)+\left( 6\underline{D} \right)\left( 2\rho -1 \right)\left( 2\rho -2 \right)\rho \right].
\label{eqA6}
\end{equation}
To make Eq. (\ref{eqA6}) hold at any $T$ and $\rho$, one can assume
\begin{subequations}
    \begin{equation}
    \left\{ \begin{aligned}
    & \frac{{\underline{A}}}{{{\kappa }_{\rho }}}=\frac{{\underline{B}}}{{{\kappa }_{\eta }}}, \\ 
    & \frac{\underline{C}+\underline{D}}{{{\kappa }_{\rho }}}=\frac{6\underline{D}}{{{\kappa }_{\eta }}}, \\ 
    \end{aligned} \right.
    \label{eqA7a}
    \end{equation}
    where
    \begin{equation*}
        \begin{aligned}
            & \underline{C}={{{\underline{C}}}_{\text{pt}}}-{{{\underline{C}}}_{\text{cf}}}\left( T-{T_\text{M}} \right), \\ 
            & \underline{D}={{{\underline{D}}}_{\text{pt}}}-{{{\underline{D}}}_{\text{cf}}}\left( T-{T_\text{M}} \right). \\ 
        \end{aligned}
    \end{equation*}
    Known that $\underline{A}+\underline{B}=1$, strong constraint above can be also rewritten as
    \begin{equation}
    \left\{ \begin{aligned}
    & \underline{A}=\frac{{{\kappa }_{\rho }}}{{{\kappa }_{\rho }}+{{\kappa }_{\eta }}},\text{      }\underline{B}=\frac{{{\kappa }_{\eta }}}{{{\kappa }_{\rho }}+{{\kappa }_{\eta }}}, \\ 
    & \frac{{{{\underline{C}}}_{\text{pt}}}+{{{\underline{D}}}_{\text{pt}}}}{{{\kappa }_{\rho }}}=\frac{6{{{\underline{D}}}_{\text{pt}}}}{{{\kappa }_{\eta }}},\frac{{{{\underline{C}}}_{\text{cf}}}+{{{\underline{D}}}_{\text{cf}}}}{{{\kappa }_{\rho }}}=\frac{6{{{\underline{D}}}_{\text{cf}}}}{{{\kappa }_{\eta }}}. \\ 
    \end{aligned} \right.
    \label{eqA7b}
    \end{equation}
\label{eqA7}
\end{subequations}
Eq. (\ref{eqA7b}) shows the relation between the model parameters $\underline{A},~\underline{B},~{{\underline{C}}_{\text{pt}}},~{{\underline{C}}_{\text{cf}}},~{{\underline{D}}_{\text{pt}}},~{{\underline{D}}_{\text{cf}}},~{{\kappa }_{\rho }}$ and ${{\kappa }_{\eta }}$. We can furthermore yield the expression ${{\gamma }_{\text{sf}}}$ for the specific surface energy. After replacing $\rho$ by $\eta$ in Eq. (\ref{eqA2}), one can get the following Euler-Lagrange equation.
\begin{equation}
\frac{\partial f\left( T,\rho ,\left\{ \rho ,0 \right\} \right)}{\partial \rho }-T\left( {{\kappa }_{\rho }}+{{\kappa }_{\eta }} \right)\nabla _{\mathbf{r}}^{2}\rho =0.
\label{eqA8}
\end{equation}
By integrating both sides of Eq. (\ref{eqA8}), one can get
\begin{equation}
f\left( T,\rho ,\left\{ \rho ,0 \right\} \right)-\frac{1}{2}T\left( {{\kappa }_{\rho }}+{{\kappa }_{\eta }} \right){{\left| {{\nabla }_{\mathbf{r}}}\rho  \right|}^{2}}={{C}_{1}},
\label{eqA9}
\end{equation}
and the arbitrary constant ${{C}_{1}}$ equals zero considering the boundary conditions in the Eq. (\ref{eqA1}). Then one can obtain the explicit formulation of the specific surface energy at equilibrium by substituting Eq. (\ref{eqA9}) into Eq. (\ref{eqA2}), i.e.
\begin{equation}
\begin{split}
{{\gamma }_{\text{sf}}}&=\int_{-\infty }^{\infty }{\left[ 2f\left( T,\rho ,\left\{ \rho ,0 \right\} \right) \right]\text{d}\mathbf{r}}, \\ 
&=\text{2}\sqrt{\frac{T\left( {{\kappa }_{\eta }}+{{\kappa }_{\rho }} \right)}{2}}\int_{0}^{1}{\sqrt{{{f}_{\text{ht}}}\rho +\left( \underline{C}+7\underline{D} \right){{\left( 1-\rho  \right)}^{2}}{{\rho }^{2}}}\text{d}\rho }. \\ 
\end{split}
\label{eqA10}
\end{equation}

Likewise, grain boundary energy can be derived. In this case, we consider two grains where the density variation is neglected. Then we have $\rho \approx 1$ across them. There are only two order parameters  ${{\eta }_{\alpha }}$ and  ${{\eta }_{\beta }}$  indicating the grains of both side. Assuming an isotropic grain boundary, profiles of ${{\eta }_{\alpha }}$ and ${{\eta }_{\beta }}$ across a grain boundary can be described by the following boundary conditions
\begin{equation}
\left\{  
\begin{array}{lr}  
{{\eta }_{\alpha }}=1,{{\eta }_{\beta }}=0  &  r\to -\infty  \\
{{\eta }_{\alpha }}=0,{{\eta }_{\beta }}=1  &  r\to +\infty  \\ 
{{\nabla }_{\mathbf{r}}}{{\eta }_{\alpha }}={{\nabla }_{\mathbf{r}}}{{\eta }_{\beta }}=0  &  r\to \pm \infty \\ 
\end{array}  
\right.
\label{eqA11}
\end{equation}
and the specific grain boundary energy can be calculated by
\begin{equation}
{{\gamma }_{\text{gb}}}=\int_{-\infty }^{\infty }{\left[ f\left( T,1,\left\{ {{\eta }_{\alpha }},{{\eta }_{\beta }} \right\} \right)+\frac{1}{2}T{{\kappa }_{\eta }}{{\left| {{\nabla }_{\mathbf{r}}}{{\eta }_{\alpha }} \right|}^{2}}+\frac{1}{2}T{{\kappa }_{\eta }}{{\left| {{\nabla }_{\mathbf{r}}}{{\eta }_{\beta }} \right|}^{2}} \right]}\text{d}\mathbf{r}.
\label{eqA12}
\end{equation}
At equilibrium, the Euler-Lagrange equation across the grain boundary reads 
\begin{equation}
\frac{\partial f\left( T,1,\left\{ {{\eta }_{\alpha }},{{\eta }_{\beta }} \right\} \right)}{\partial {{\eta }_{\alpha }}}-T{{\kappa }_{\eta }}\nabla _{\mathbf{r}}^{2}{{\eta }_{\alpha }}=\frac{\partial f\left( T,1,\left\{ {{\eta }_{\alpha }},{{\eta }_{\beta }} \right\} \right)}{\partial {{\eta }_{\beta }}}-T{{\kappa }_{\eta }}\nabla _{\mathbf{r}}^{2}{{\eta }_{\beta }}=0 .
\label{eqA14}
\end{equation}
Since we constrain the summation of all non-conserved ${{\eta }_{\alpha }}$ to be unity, i.e. ${{\eta }_{\beta }}=1-{{\eta }_{\alpha }}$, it is easy to find that ${{\nabla }_{\mathbf{r}}}{{\eta }_{\alpha }}=-{{\nabla }_{\mathbf{r}}}{{\eta }_{\beta }}$. We can thereby take a single order parameter $\eta $ to replace ${{\eta }_{\alpha }}$ and ${{\eta }_{\beta }}$ in Eq. (\ref{eqA12}) by setting ${{\eta }_{\alpha }}=\eta $ and ${{\eta }_{\beta }}=1-\eta $. Following Eq. (\ref{eqA14}), it yields
\begin{equation}
\frac{\partial f\left( T,1,\left\{ \eta ,1-\eta  \right\} \right)}{\partial \eta }-2T{{\kappa }_{\eta }}\nabla _{\mathbf{r}}^{2}\eta =0.
\label{eqA15}
\end{equation}
By integrating both sides of Eq. (\ref{eqA15}) one can get
\begin{equation}
f\left( T,1,\left\{ {{\eta }_{\alpha }},{{\eta }_{\beta }} \right\} \right)-T{{\kappa }_{\eta }}{{\left| {{\nabla }_{\mathbf{r}}}\eta  \right|}^{2}}={{C}_{2}},
\label{eqA16}
\end{equation}
and the arbitrary constant ${{C}_{2}}$ equals zero considering the boundary conditions in  Eq. (\ref{eqA11}). Substituting Eq. (\ref{eqA16}) into Eq. (\ref{eqA12}), we eventually obtain the explicit formulation of the specific grain-boundary energy at equilibrium
\begin{equation}
\begin{split}
{{\gamma }_{\text{gb}}}&=\int_{-\infty }^{\infty }{\left[ 2f\left( T,1,\left\{ \eta ,1-\eta  \right\} \right) \right]\text{d}\mathbf{r}}, \\ 
&=\text{2}\sqrt{T{{\kappa }_{\eta }}}\int_{0}^{1}{\sqrt{{{f}_{\text{ht}}}+12\underline{D}{{\left( 1-\eta  \right)}^{2}}{{\eta }^{2}}}\text{d}\eta }. 
\end{split}
\label{eqA17}
\end{equation}

\begin{figure*}[!t]
\centering
\includegraphics[width=10cm]{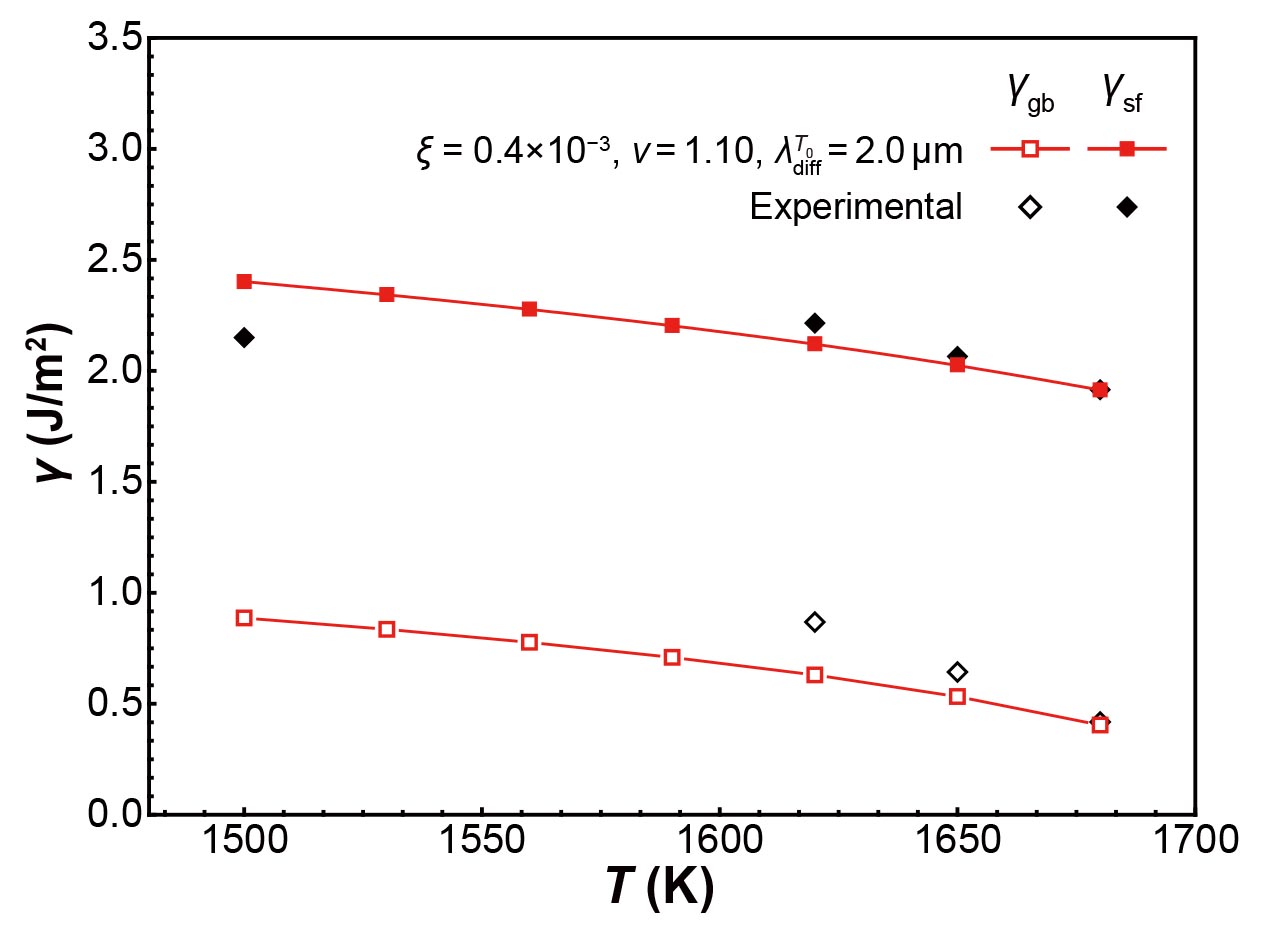}
\caption{Fitting model parameters with the temperature-dependent ${{\gamma }_{\text{sf}}}$ and ${{\gamma }_{\text{gb}}}$.}
\label{figA2}
\end{figure*}

We regard Eq. (\ref{eqA10}) and (\ref{eqA17}) as the relations which link the model parameters $\underline{A}$, $\underline{B}$, $\ {{\underline{C}}_{\text{pt}}}$, $\ {{\underline{C}}_{\text{cf}}}$, $\ {{\underline{D}}_{\text{pt}}}$, ${{\underline{D}}_{\text{cf}}}$, ${{\kappa }_{\eta }}$ and ${{\kappa }_{\rho }}$ to the measurable, temperature-dependent material properties ${{\gamma }_{\text{sf}}}\left( T \right)$ and ${{\gamma }_{\text{gb}}}\left( T \right)$, which are usually linearly fitted in practical measurement. Due to the dependency in Eq. \ref{eqA7b}, there are only four independent quantities in these eight model parameters, i.e. ${{\underline{D}}_{\text{pt}}}\text{, }{{\underline{D}}_{\text{cf}}},{{\kappa }_{\rho }},{{\kappa }_{\eta }}$ (or ${{\underline{C}}_{\text{pt}}}\text{, }{{\underline{C}}_{\text{cf}}},{{\kappa }_{\rho }},{{\kappa }_{\eta }}$). Here we present a method utilizing fitting parameters to fit the temperature-dependent ${{\gamma }_{\text{sf}}}$ and ${{\gamma }_{\text{gb}}}$ to avoid the direct solving by formulating 4 equations based on Eqs. (\ref{eqA10}) and (\ref{eqA17}). This method firstly relates the ${{\kappa }_{\rho }},{{\kappa }_{\eta }}$ and ${{\underline{C}}_{\text{pt}}}$, ${{\underline{D}}_{\text{pt}}}$ to the experimental values of the surface and grain-boundary energies approaching the melting temperature $T_\text{M}$, i.e.
\begin{equation}
\begin{split}
\gamma _{\text{sf}}^{{T_\text{M}}} &=\text{2}\sqrt{\frac{{T_\text{M}}\left( {{\kappa }_{\eta }}+{{\kappa }_{\rho }} \right)}{2}}\int_{0}^{1}{\sqrt{\left( {{{\underline{C}}}_{\text{pt}}}+7{{{\underline{D}}}_{\text{pt}}} \right){{\left( 1-\rho  \right)}^{2}}{{\rho }^{2}}}\text{d}\rho } \\ 
&=\frac{1}{3\sqrt{2}}\sqrt{{T_\text{M}}\left( {{\kappa }_{\eta }}+{{\kappa }_{\rho }} \right)\left( {{{\underline{C}}}_{\text{pt}}}+7{{{\underline{D}}}_{\text{pt}}} \right)}, \\ 
\gamma _{\text{gb}}^{{T_\text{M}}}&=\text{2}\sqrt{{T_\text{M}}{{\kappa }_{\eta }}}\int_{0}^{1}{\sqrt{12{{{\underline{D}}}_{\text{pt}}}{{\left( 1-\eta  \right)}^{2}}{{\eta }^{2}}}\text{d}\eta } \\ 
&=\frac{2}{\sqrt{3}}\sqrt{{T_\text{M}}{{\kappa }_{\eta }}{{{\underline{D}}}_{\text{pt}}}}, \\ 
\end{split}
\label{eqA18}
\end{equation}
where ${{\underline{C}}_{\text{pt}}}$ and ${{\underline{D}}_{\text{pt}}}$ are related according to Eq. (\ref{eqA7b}). In the phase-field model, the grain boundary is a diffusive interface, and its approximate width ${{\lambda }_{\text{gb}}}$ at $T_{0}$ can be expressed as
\begin{equation}
{{\lambda}^{T_{0}}_{\text{gb}}}\approx \frac{1}{\tan \left( \theta^{T_{0}} /2 \right)}=\frac{1}{{{\left. {{\nabla }_{\mathbf{r}}}\eta^{T_{0}}  \right|}_{{{r}_{0}}}}} = \sqrt{4{T_\text{M}}{{\kappa }_{\eta }}/3{{{\underline{D}}}_{\text{pt}}}} .
\label{eqA19}
\end{equation}
Based on Eq. (\ref{eqA7b}), at $T_\text{M}$ we can define a fitting parameter $\nu$ as
\begin{equation}
\nu ={T_\text{M}}{{\underline{D}}_{\text{cf}}}/{{\underline{D}}_{\text{pt}}}={T_\text{M}}{{\underline{C}}_{\text{cf}}}/{{\underline{C}}_{\text{pt}}}
\label{eqA19-2}
\end{equation}
By using the four Eqs. (\ref{eqA18})--(\ref{eqA19-2}) at $T_\text{M}$, we can determine $\underline{D}_\text{pt}$, $\underline{D}_\text{cf}$, $\kappa_\rho$, $\kappa_\eta$ and thus eight model parameters according to the dependency. Then according to Eqs. (\ref{eqA10}) and (\ref{eqA17}), the temperature-dependent $\gamma_\text{sf} \left(T \right)$ and $\gamma_\text{gb}\left( T \right)$ can be obtained. 

Normally we can fit ${{\gamma }_{\text{sf}}}\left( T \right)$ and ${{\gamma }_{\text{gb}}}\left( T \right)$ to make them close to the experimental data by giving ${\lambda }_{\text{gb}}^{T_\text{M}}$ and tuning $\nu$. However, it is too thin for the real grain boundary width (usually several nanometers) when compare to the size of the particles/grains in the scale of micrometer, requiring much finer mesh in finite element simulations \citep{yang2018arxiv}. To deal with such a problem, a common solution is to manually set the diffusion width ${\lambda }_{\text{gb}}^{T_\text{M}}$ close to the scale of the particles/grains, e.g. $1\sim2$ \si{\micro\meter} for the particles/grains with an average size of several tens micrometer, while keep the $\gamma_\text{gb}$ identical. According to Eqs. (\ref{eqA18}) and (\ref{eqA19}), the barrier height of the potential part of the free energy density $f_\text{pt}=12\underline{D}\left(1-\eta\right)^2\eta^2$, which is proportional to the $\underline{D}$, will decrease when wider $\gamma_\text{gb}$ is utilized to keep $\gamma_\text{gb}$ identical. In this case, the heat part of the free energy density $f_\text{ht}$ in Eqs. (\ref{eqA10}) and (\ref{eqA17}) should be modified as $\xi f_\text{ht}$ by introducing a coefficient $\xi$ to have the same scale as the potential part $f_\text{pt}$. Fig. \ref{figA2}a shows the fitted $\gamma_\text{sf}$ and $\gamma_\text{gb}$ by setting ${\lambda }^{T_{\text{M}}}_{\text{gb}}=2~\si{\micro\meter}$ ($T_\text{M}=1700$ K), $\xi=0.3\times10^{-3}$, and $\nu=1.01$, as well as the experimental $\gamma_\text{sf}$ and $\gamma_\text{gb}$. 
\newpage


\bibliography{sup_reference}
